\documentclass[a4paper,11pt]{article}

\usepackage{amsmath,amsfonts,amssymb,caption,graphicx,natbib}
\input{epsf}

\def\tiid{\text{iid}}
\def\trev{\text{rev}}

\newcommand{\weakly}{\mbox{$ \;\stackrel{\cal D}{\longrightarrow}\; $}}

\newcommand{\eqquestion}{\mbox{$ \;\stackrel{?}{=}\; $}}

\newcommand{\RL}{{\mathbb R}}

\newcommand{\IND}{{\mathbb I}}

\newcommand{\PR}{\mbox{\rm Pr}} 
\newcommand{\K}{{\rm K}}
 
\newcommand{\VAR}{\mbox{\rm Var}}

\newcommand{\COV}{\mbox{\rm Cov}}

\newcommand{\iid}{\mbox{i.i.d.}\!}

\def\be{\begin{eqnarray}}
\def\ee{\end{eqnarray}}
\def\ben{\begin{eqnarray*}}
\def\een{\end{eqnarray*}}

\def\slabel#1{\label{s:#1}}
\def\flabel#1{\label{f:#1}}
\def\elabel#1{\label{e:#1}}

\def\sq{$\Box$}

\def\qed{\ifmmode\sq\else{\unskip\nobreak\hfil
\penalty50\hskip1em\null\nobreak\hfil\sq
\parfillskip=0pt\finalhyphendemerits=0\endgraf}\fi\par\medbreak}

\newsavebox{\junk}
\savebox{\junk}[1.6mm]{\hbox{$|\!|\!|$}}

\def\state{{\sf X}}

\def\til={{\widetilde =}}

\def\clB{{\cal B}}

\def\half{{\mathchoice{\textstyle \frac{1}{2}}%
{\frac{1}{2}}%
{\hbox{\tiny $\frac{1}{2}$}}%
{\hbox{\tiny $\frac{1}{2}$}} }}

 \def\eq#1/{(\ref{#1})}

\def\Section#1{Section~\ref{s:#1}}

\def\Figure#1{Figure~\ref{f:#1}}

\def\eq#1/{(\ref{e:#1})}

\newcommand{\beqn}[1]{\notes{#1}%
\begin{eqnarray} \elabel{#1}}

\newcommand{\eeqn}{\end{eqnarray} }

\newcommand{\beq}[1]{\notes{#1}%
\begin{equation}\elabel{#1}}

\newcommand{\eeq}{\end{equation}} 

\def\bdes{\begin{description}}
\def\edes{\end{description}}

\def\notes#1{}

\begin{document}

\title{\vspace{-1.5cm}%
Notes on Using Control Variates\\
for Estimation with Reversible MCMC Samplers}

\author
{
    Petros Dellaportas
    \thanks{Department of Statistics,
        Athens University of Economics and Business,
        Patission 76, Athens 10434, Greece.
                Email: {\tt petros@aueb.gr}.
        }
\and
        Ioannis Kontoyiannis
    \thanks{Department of Informatics,
        Athens University of Economics and Business,
        Patission 76, Athens 10434, Greece.
                Email: {\tt yiannis@aueb.gr}.
        }
}

\date{July 2009}

\maketitle

\begin{abstract}
A general methodology is presented for the construction and
effective use of control variates for reversible MCMC samplers. 
The values of the coefficients of the optimal linear combination 
of the control variates are computed, and adaptive, consistent 
MCMC estimators are derived for these optimal coefficients. 
All methodological and asymptotic arguments are rigorously justified.  
Numerous MCMC simulation examples from Bayesian inference 
applications demonstrate that the resulting variance reduction 
can be quite dramatic.
\end{abstract}



\thispagestyle{empty}
\setcounter{page}{1}

\section{Introduction}
\slabel{intro}

Markov chain Monte Carlo (MCMC) methods provide the facility to
draw, in an asymptotic sense, a sequence of dependent samples
from a very wide class of probability measures in any dimension. 
This facility -- together with the tremendous 
increase of computer power in recent years -- makes MCMC 
perhaps the main reason for the widespread use of Bayesian 
statistical modeling across the entire spectrum 
of quantitative scientific disciplines. 

This paper provides a firm methodological foundation for
the construction and use of control variates for reversible 
MCMC samplers. 
Although popular in the standard Monte Carlo setting, 
control variates have received little attention
in the MCMC literature.
The proposed methodology will be shown, in many instances,
to reduce the variance of the resulting estimators 
quite dramatically.

In the simplest Monte Carlo setting, when
the goal is to compute the expected value 
of some function $F$ evaluated on independent 
and identically distributed
(i.i.d.) samples $X_1,X_2,\ldots$, the variance 
of the standard ergodic averages of the $F(X_i)$
can be reduced by exploiting available zero-mean 
statistics. If there are one or more functions 
$U_1,U_2,\ldots,U_k$ --
the {\em control variates} -- for which it
is known that the expected value of $U(X_i)$
is equal to zero, then adding any linear 
combination, $\theta_1U_1(X_i)+
\theta_2U_2(X_i)+\cdots+
\theta_kU_k(X_i),$
to the $F(X_i)$ does not change 
the asymptotic mean of the corresponding
ergodic averages. 
Moreover,
if the best constant coefficients $\{\theta^*_j\}$ 
are used, then the variance
of the estimates is no larger than before and
often it is much smaller. The standard practice
in this setting is to estimate the
optimal $\{\theta_j^*\}$ adaptively, based on the
same sequence of samples;
see, e.g., \citet{liu:book}, \citet{givens:book},
\citet{robert:book} for details. 
Because of the demonstrated effectiveness
of this technique, in many important areas
of application -- e.g., in computational
finance where Monte Carlo methods are a basic
tool for the approximate computation 
of expectations, see \cite{glasserman:book} --
a major research effort is devoted to
the construction of effective control variates
in specific applied problems.

However, up to now little has been established in the way
of extending the above methodology to
estimators based on MCMC samples, at least in part due to 
the intrinsic difficulties presented by the Markovian
structure. For example, 
\citet{mr1} comment
that ``control variates have been advertised early in the MCMC
literature (see, e.g., \citet{greenhan}), but they 
are difficult to work 
with because the models are
always different and their complexity is such that it is
extremely challenging to
derive a function with known expectation.''
Indeed, there are two fundamental difficulties;
not only is it hard to find nontrivial functions
with known expectation with respect to the stationary
distribution of the chain, but also, even in cases
where such functions are available, there is no
effective way to obtain useful estimates of the
corresponding optimal coefficients $\{\theta_j^*\}$.
The reason
why this is a fundamentally difficult problem
is that the MCMC variance of ergodic averages is
intrinsically an infinite-dimensional object:
It cannot be written in closed form as a function 
of the transition kernel and the stationary distribution
of the chain.

An early reference of variance reduction for Markov chain samplers
is \citet{greenhan}, who exploit an idea of \citet{barfri89}
and construct antithetic variables that may achieve variance
reduction in simple settings but they do not appear to
be widely applicable.
\citet{adra} focus on finite state space chains,
they observe that optimum variance reduction can be
achieved via the solution of the associated 
Poisson equation
(see \Section{general}), and they propose numerical algorithms 
for its solution.
Rao-Blackwellisation has been suggested by \cite{gelfand+smith} and
by \citet{robert:book} as a way to reduce the variance of MCMC
estimators. Also, \citet{Philrob} investigated the use of Riemann 
sums as a variance reduction tool in MCMC algorithms. An interesting as
well as natural control variate that has been used, mainly as
a convergence diagnostic, by \citet{FanBrGe}, is the score statistic.
Although \citet{Philrob} mention that it can be used as a control
variate, its practical utility has not been investigated.
\citet{Atchade_perron} restrict attention to independent Metropolis
samplers and provide an explicit formula 
for the construction of control variates based
on adaptive estimators. \citet{Mira_tecrep} note that a solution to
the Poisson equation provides the optimum control variate and they
attempt to solve it numerically. \citet{Hammer08} construct control
variates for general Metropolis-Hastings samplers by expanding the
state space. To estimate the optimal coefficients 
$\{\theta_j^*\}$ they use the same formula that one
obtains for control variates in i.i.d.\ Monte Carlo sampling,
but such estimators are strictly suboptimal; they are 
briefly discussed in Section~\ref{s:suboptimal}, where
we also explore their efficiency.

A more relevant, for our purposes, line of work is
that initiated by \cite{henderson:phd},
who observed that, for {\em any} real-valued 
function $G$ defined on the state space of 
a Markov chain $\{X_n\}$, the function 
$U(x):=G(x)-E[G(X_{n+1})|X_n=x]$
has zero mean with respect to the stationary 
distribution of the chain. 
\citet{henderson:phd}, like some of the other
authors mentioned above, also notes that the
best choice for the function $G$ would
be the solution of the associated 
Poisson equation, and proceeds to compute
approximations of this solution for specific
Markov chains, with particular emphasis
on models arising in stochastic network
theory.

The gist of our approach is to adapt Henderson's
idea for the construction of control variates,
and use them in conjunction with a new,
efficiently implementable and provably optimal
estimator for the coefficients $\{\theta_j^*\}$
for reversible chains. The ability to estimate
the $\{\theta_j^*\}$ effectively makes these
control variates practically relevant in
the statistical MCMC context, and allows 
us to avoid having
to compute analytical approximations to the
solution of the underlying Poisson equation.
Our estimator for $\{\theta_j^*\}$ is adaptive, 
in the sense that 
is based on the MCMC output, and it can be used
after the sample is obtained, making its actual 
computation independent of the MCMC algorithm.

This methodology not only generalizes the classical
method of control variates to the MCMC setting,
it also offers an important advantage:
Unlike the case of independent sampling where 
control variates need to be found in an 
{\em ad hoc} manner depending on the specific
problem at hand, here the control variates
(as well as estimates for the corresponding
optimal coefficients) come for
free. The only requirement for the application
of this method at the post-processing stage
is the availability of a function $G$ of the sampled
parameters, together with its one-step conditional
expectation, $E[G(X_{n+1})|X_n=x]$.
As we show in numerous specific examples, 
these are often readily available;
for example, the availability of such
expectations is essentially a prerequisite 
for Gibbs sampling. 

For any one particular application, there is,
of course, a plethora of functions $G$ 
(and, consequently, of corresponding control variates $U$)
that can be used, so an important consideration for the 
effectiveness of this methodology
for variance reduction is the careful
choice of these functions.
This issue is addressed in detail; 
we provide numerous illustrative examples
of estimation problems based on MCMC samplers,
motivated primarily by Bayesian inference
problems. These examples are chosen as
representing different major classes of
MCMC samplers commonly used in important
applications. In each case, the ideas
underlying the choice of the functions $G$ 
are explained, and these choices are
justified either rigorously or heuristically,
in connection with the theoretical development
we present.

The examples we consider range from the simplest,
illustrative samplers, to complex
applications of Bayesian models to real data.
In all cases, the resulting variance reduction 
is very significant and often quite large:
For all the MCMC-based ergodic estimators we consider,
the use of control variates gives 
variances at least $30$ times smaller,
and often hundreds or thousands of times
smaller.

Presently we focus only on cases of reversible
MCMC samplers for which the one-step
conditional expectations, 
$E[G(X_{n+1})|X_n=x]$,
of one or more functions $G$ are
available analytically in closed form.
MCMC algorithms with this property include a 
vast array of samplers commonly used in 
practical Bayesian inference problems.  
In the examples presented in Sections~\ref{sec:simple}
and~\ref{s:complex} below we outline the implementation details
of our methodology for a representative subset of both simple 
and complex models.  Since our estimators for 
$\{\theta_j^*\}$ are applicable to reversible chains, 
we employ random-scan instead of the usual 
systematic-scan Gibbs or 
Metropolis-within-Gibbs algorithms. We also investigate 
the behavior of our estimators on discrete state space,
random-walk Metropolis-Hastings samplers, 
and on Metropolis-within-Gibbs samplers.  
Although, strictly speaking, our theoretical development 
does not necessarily require that conditional expectations
$E[G(X_{n+1})|X_n=x]$ be
analytically available, almost all of the examples 
presented here do have that property, primarily for the 
sake of convenience and of clarity of exposition.
Further ongoing work by \citet{DKMT:prep} explores 
ways in which this same theory can be applied
to arbitrary reversible MCMC samplers, including
cases where one-step conditional expectations
are unavailable.

As mentioned above, \cite{henderson:phd}
takes a different path toward optimizing
the use of control variates for Markov chain
samplers. Considering primarily continuous-time 
Markov processes, an approximation $G$ for the 
solution to the associated Poisson equation is 
derived from the so-called ``heavy traffic'' 
or ``fluid model'' approximations of the 
original process. The motivation and application
of this method is primarily related to examples
from stochastic networks and queueing theory.
Closely related approaches are presented
by \cite{henderson-glynn:02} and
\cite{HMT:03}, where the effectiveness of
network control policies of multiclass networks
is evaluated via Markovian simulation tools.
There, control variates are used for variance 
reduction, and the optimal parameters $\{\theta_j^*\}$
are estimated via an adaptive, stochastic gradient algorithm. 
General convergence properties of ergodic estimators
using control variates are derived by \cite{henderson-simon:04},
in the case when the solution to the Poisson equation 
(either for the original chain or for an approximating
chain) is known explicitly.
\cite{kim-henderson:07} introduce two related 
adaptive methods
for tuning non-linear versions of the parameters
$\{\theta_j\}$, when using families of control
variates that naturally admit a non-linear parameterization.
After deriving asymptotic properties for these
estimators, they present numerical examples for 
a simulation problem related to pricing derivative
instruments in computational finance.
In the case when the control variate $U$ is
defined in terms of a function $G$ that can 
be taken as a Lyapunov function for the 
chain $\{X_n\}$, \cite{meyn:06} derives 
precise exponential asymptotics for the
performance of estimators employing such
control variates.

The rest of the paper is organized as follows.  
\Section{general} gives 
the basic definitions that will remain in effect throughout the paper,
and motivates the construction of control variates in connection
with the Poisson equation.
Sections~\ref{s:cv2}, \ref{s:suboptimal} and~\ref{s:theta-rev}, 
building on ideas
of \citet{henderson:phd}, illustrate the use of naive estimators 
of the optimal coefficient for a single control variate, 
and develop the theory for two new estimators 
for reversible chains.  In Section~3 we investigate the impact 
of these estimators on variance reduction in five small 
MCMC examples, which are
representative of a larger class of Bayesian inference problems.
Section~4 discusses the effect of our estimators on bias reduction,
compares the two estimators and advocates the use of one of them for
general purposes.  These estimators are generalized in Section~5 to
the case of multiple control variates.  Four more complex Bayesian
inference problems that are implemented via MCMC are visited in
Section~6; guidelines for constructing appropriate control variates
are given, and their effects on variance reduction are illustrated.  
Finally, we provide theoretical justifications of our asymptotic 
arguments in Section~7 and conclude with a short discussion 
of possible further extensions in Section~8.


\section{Control Variates for Markov Chains}
\slabel{cv}

\subsection{The setting}
\slabel{general}

Suppose $\{X_n\}$ is a discrete-time Markov chain with initial state
$X_0=x$, taking values in the state space $\state$ with an
associated $\sigma$-algebra $\clB$. In typical applications,
$\state$ will often be a (Borel measurable) subset of $\RL^d$
together the collection $\clB$ of all its (Borel) measurable
subsets. [More precise definitions and detailed assumptions will be
given in \Section{theory}.] The distribution of $\{X_n\}$ is
described by its transition kernel, $P(x,dy)$,
\be
P(x,A):=\Pr\{X_{k+1}\in A\,|\,X_k=x\},
\;\;\;\;x\in\state,\;A\in\clB. \label{eq:kernel} 
\ee

It is well known that in many applications where
it is desirable to compute the expectation
$E_\pi(F):=\pi(F):=\int F\,d\pi$
of some function $F:\state\to\RL$ with respect to some
probability measure $\pi$ on $(\state,\clB)$,
it turns out that, although the direct computation
of $\pi(F)$ is impossible or we cannot even produce
samples from $\pi$, we can construct an
easy-to-simulate Markov chain $\{X_n\}$
which has $\pi$ as its unique invariant measure.
Under appropriate conditions, the distribution
of $\{X_n\}$ converges to $\pi$, a fact which
can be made
precise in several ways. For example,
writing
$PF$ for the function,
$$PF(x):=E_x[F(X_1)]:=E[F(X_1)\,|\,X_0=x],
\;\;\;\;x\in\state,$$
we have that, for any initial condition $x$,
$$P^nF(x):=E[F(X_n)\,|\,X_0=x]\to\pi(F),\;\;\;\;\mbox{as}\;n\to\infty,$$
for an appropriate class of functions $F:\state\to\RL$. Furthermore,
the rate of this convergence can be quantified by the function, \be
\hat{F}(x)=\sum_{n=0}^\infty \Big [P^nF(x)-\pi(F)\Big],
\label{eq:sumFhat} \ee where $\hat{F}$ is easily seen to satisfy the
Poisson equation for $F$, namely, \be P\hat{F}-\hat{F}=-F+\pi(F).
\label{eq:poisson} \ee To see that, at least formally, simply apply
$P$ to both sides of (\ref{eq:sumFhat}) and note that the resulting
series for $P\hat{F}-\hat{F}$ becomes telescoping and simplifies to
$-F+\pi(F)$.

The above results describe how the distribution
of $X_n$ converges to $\pi$. In terms of estimation,
the quantities of interest are the ergodic averages,
$$\mu_n(F):=\frac{1}{n}\sum_{i=0}^{n-1}F(X_i).$$
Again, under appropriate conditions the ergodic theorem
holds,
\be
\mu_n(F)\to\pi(F),\;\;\;\;\mbox{a.s., as}\;n\to\infty,
\label{eq:ergodic}
\ee
for an appropriate class of functions $F$.
Moreover, the rate of this convergence is quantified
by an associated central limit theorem, which states
that,
$$\sqrt{n}[\mu_n(F)-\pi(F)]=
\frac{1}{\sqrt{n}}\sum_{i=0}^{n-1}[F(X_i)-\pi(F)] \weakly
N(0,\sigma_F^2), \;\;\;\;\mbox{as}\;n\to\infty,$$ where
$\sigma_F^2$, the asymptotic variance of $F$, is given by,
$$\sigma_F^2:=\lim_{n\to\infty}\VAR_\pi(\sqrt{n}\mu_n(F))
=\lim_{n\to\infty}\VAR_\pi\Big(\frac{1}{\sqrt{n}}
\sum_{i=0}^{n-1}F(X_i)\Big)
=
\sum_{n=-\infty}^\infty
\COV_\pi(F(X_0),F(X_n)).$$
Alternatively, it can be expressed in terms of the
solution $\hat{F}$ to Poisson's equation as,
\be
\sigma_F^2=\pi\Big(\hat{F}^2-(P\hat{F})^2\Big).
\label{eq:varF}
\ee

The results in equations (\ref{eq:sumFhat})
and (\ref{eq:varF}) clearly indicate that
it is
useful to be able to compute the solution
$\hat{F}$ to the Poisson equation for $F$.
In general this is a highly nontrivial -- or impossible --
task;
for one thing, it requires knowledge of the
mean $\pi(F)$. The following example is one
of the rare cases where explicit computations are possible.

Suppose $\{X_n\}$ is a discrete time version
of the Ornstein-Uhlenbeck process defined by,
$X_0=x$ and $X_{n+1}=\alpha X_n + Z_n,$
where $\alpha$ is a constant in $(0,1)$ and
$\{Z_n\}$ are independent and identically
distributed ($\iid$) standard Normal
random variables.
Standard methods easily show that the
distribution of $X_n$ converges
to $\pi:=N(0,(1-\alpha^2)^{-1})$,
so if we take $F(x)\equiv x$,
then, $\mu_n(F)\to\pi(F)=0$ a.s., as $n\to\infty$.
Moreover, the central limit theorem implies that,
$$\sqrt{n}\mu_n(F)\weakly N(0,\sigma^2),
\;\;\;\;n\to\infty,$$
where $\sigma^2=\sigma_F^2$ is given by (\ref{eq:varF}).
In order to compute the variance
we need to know $\hat{F}$. As a first
guess, we take $G(x)=cx+b$ and compute,
$$PG(x)-G(x)
=E[cX_1+b\,|\,X_0=x]-cx-b
=E[c(\alpha x+Z_1)]-cx
=-c(1-\alpha)x.$$
For this to be equal to $-F(x)+\pi(F)=-x$, we need
$c=(1-\alpha)^{-1}$; any $b$ will do. Taking, for simplicity, $b=0$,
yields,
$$
\hat{F}(x)=\frac{x}{1-\alpha}
\;\;\;\;
\mbox{and}
\;\;\;\;
P\hat{F}(x)=\frac{\alpha x}{1-\alpha},
\;\;\;\;
x\in\RL.$$
Therefore, writing $W$ for
$N(0,(1-\alpha^2)^{-1})$ random variable,
\ben
\sigma^2
=
    \pi\Big(\hat{F}^2-(P\hat{F})^2\Big)
=
    E\Big[\frac{W^2}{(1-\alpha)^2}
    -
    \frac{\alpha^2W^2}{(1-\alpha)^2}
    \Big]
=
    \frac{1}{(1-\alpha)^2}.
\een


\subsection{Control variates}
\slabel{cv2}

Suppose that, for some Markov chain
$\{X_n\}$ with transition kernel $P$ and invariant
measure $\pi$, we use the ergodic averages
$\mu_n(F)$ as in (\ref{eq:ergodic})
to estimate the mean $\pi(F)$
of some function $F$
under $\pi$. In many applications,
although the estimates $\mu_n(F)$
converge to $\pi(F)$ as $n\to\infty$,
the associated asymptotic variance $\sigma_F^2$
is large and the
convergence is very slow.

In order to reduce the variance, we employ the idea of using control
variates, as in the case of simple Monte Carlo with $\iid$ samples;
see, for example, the standard texts
\citet{robert:book,liu:book,givens:book}, or the paper by
\citet{glynn-szechtman} for extensive discussions. Given a
function $U:\state\to\RL$ for which we know that $\pi(U)=0$, define,
\be F_\theta=F-\theta U, \label{eq:Ftheta} \ee and consider the
modified estimators,
$$\mu_n(F_\theta)=\mu_n(F)-\theta\mu_n(U).$$
We will concentrate exclusively on the
the following class of functions $U$ proposed by
\citet{henderson:phd}. For an arbitrary $G:\state\to\RL$
with $\pi(|G|)<\infty$, define,
$$U=G-PG.$$
The invariance of $\pi$ under $P$ and  the integrability
of $G$ immediately imply that $\pi(U)=0$.
[See \Section{theory} for the
details, complete assumptions, and
full, rigorous results corresponding
to this discussion.]
Therefore, the ergodic theorem guarantees
that the $\{\mu_n(F_\theta)\}$ are consistent with probability
one, and it is natural to seek particular choices
for $U$ and $\theta$ so that the asymptotic
variance $\sigma_{F_\theta}^2$ of the modified
estimators
is significantly smaller that the variance $\sigma_F^2$
of the standard ergodic averages $\mu_n(F)$.

Suppose, at first, that we have complete freedom
in the choice of $G$, so that we may set $\theta=1$
without loss of generality. Then
we wish to make the asymptotic variance of,
$$F-U=F-G+PG,$$
as small as possible. But, in view of the Poisson equation
(\ref{eq:poisson}), we see that the choice $G=\hat{F}$ yields, \be
F-U=F-\hat{F}+P\hat{F}=\pi(F), \label{eq:zero} \ee which has zero
variance. Therefore, our first rule of thumb for choosing $G$ is:

\begin{center}
{\em Choose a control variate $U=G-PG$ with
$G\approx\hat{F}$}.
\end{center}

\noindent
As mentioned above, it is typically impossible
to compute $\hat{F}$ for realistic models in
applications. But it is often possible to come
up with a guess $G$ that approximates $\hat{F}$,
or at least some $G$ for which heuristics
indicate that it would be useful as a control
variate. Once such a function is selected,
we form the modified estimators $\mu_n(F_\theta)$
with respect to the function $F_\theta$ as in
(\ref{eq:Ftheta}),
$$F_\theta=F-\theta U=F-\theta G+\theta PG.$$

The next task is to choose $\theta$ so that the
resulting variance,
$$\sigma_\theta^2:=\sigma_{F_\theta}^2=
\pi\Big(\hat{F}_\theta^2-(P\hat{F}_\theta)^2\Big),$$
is minimized. Note that, from the definitions,
\be
\hat{U}=G\;\;\;\;\mbox{and}
\;\;\;\;
\hat{F}_\theta=\hat{F}-\theta G.
\label{eq:FthetaHat}
\ee
Therefore,
$$\sigma_\theta^2=
\pi\Big((\hat{F}-\theta G)^2\Big)-\pi\Big((P\hat{F}-\theta PG)^2\Big).$$
Expanding the above expression as a quadratic in $\theta$,
the optimal value for $\theta$ is determined as,
\be
\theta^*=\frac
{\pi\big(\hat{F}G-(P\hat{F})(PG)\big)}
{\pi(G^2-(PG)^2)}.
\label{eq:thetaStar}
\ee
Note that, since $\hat{U}=G$, the denominator is simply
$\sigma^2_U$.
Once again, this expression depends on $\hat{F}$,
so it is not immediately clear how to estimate
$\theta^*$ directly from the data $\{X_n\}$.
We consider the issue of estimating $\theta^*$ in detail
below, but first let us interpret the optimal value
of $\theta^*$. Starting from the expression,
$$\sigma_\theta^2=\lim_{n\to\infty}
\VAR_\pi\Big(\frac{1}{\sqrt{n}}\sum_{i=0}^{n-1}[F(X_i)-\theta U(X_i)]\Big),$$
simple calculations lead to,
$$\sigma_\theta^2=\sigma_F^2+\theta^2\sigma_U^2
-2\theta\sum_{n=-\infty}^\infty\COV_\pi(F(X_0),U(X_n)),$$
so that $\theta^*$ can also be expressed as
\be
\theta^*=
\frac{1}{\sigma_U^2}
\sum_{n=-\infty}^\infty\COV_\pi(F(X_0),U(X_n)),
\label{eq:thetaRatio}
\ee
leading to the optimal asymptotic variance,
\be
\sigma_{\theta^*}^2=\sigma_F^2
-\frac{1}{\sigma_U^2}
\Big[\sum_{n=-\infty}^\infty\COV_\pi(F(X_0),U(X_n))\Big]^2.
\label{eq:thetaStar2}
\ee
Therefore, in order to reduce the variance, we want
to have the covariance between $F$ and $U$ to be
as large as possible. This leads to our second
rule of thumb for selecting control variates:

\begin{center}
{\em Choose a control variate $U=G-PG$ so
that $U$ and $F$ are highly correlated.}
\end{center}

\noindent
Incidently, note that,
since the denominator of
(\ref{eq:thetaStar}) equals
$\sigma^2_U$,
comparing the expressions
for $\theta^*$ in (\ref{eq:thetaStar}) and
(\ref{eq:thetaStar2}) we see that,
\be
\sum_{n=-\infty}^\infty\COV_\pi(F(X_0),U(X_n))
=
    \pi\big(\hat{F}G-(P\hat{F})(PG)\big).
\label{eq:cov}
\ee
Moreover, the fact that $\sigma_U^2$ is
always nonnegative, suggests that there should be
a way to rewrite the expression $\pi(G^2-(PG)^2)$
in the denominator of $\theta^*$ in a way which makes
this nonnegativity obvious. Indeed:

\medskip

\noindent
{\bf Lemma 1. } The asymptotic variance $\sigma_U^2$ of
the function $U=G-PG$ can be expressed as,
\be
\sigma_U^2 = \pi(G^2-(PG)^2) =
    E_\pi\Big[\Big(G(X_1)-PG(X_0)\Big)^2\Big].
\label{eq:lemma1}
\ee

\noindent
{\sc Proof. }
Starting from the right-hand side of (\ref{eq:lemma1}),
\ben
E_\pi\Big[\Big(G(X_1)-PG(X_0)\Big)^2\Big]
&=&
    \pi(G^2)
    -2E_\pi[G(X_1)PG(X_0)]
    +\pi((PG)^2)\\
&=&
    \pi(G^2)
    -2E_\pi\Big\{E\Big[G(X_1)PG(X_0)\,\Big|\,X_0\Big]\Big\}
    +\pi((PG)^2)\\
&=&
    \pi(G^2)
    -2E_\pi\Big[E[G(X_1)\,|\,X_0]PG(X_0)\Big]
    +\pi((PG)^2)\\
&=&
    \pi(G^2-(PG)^2),
\een
as claimed. The fact that
$\sigma_U^2 = \pi(G^2-(PG)^2)$
is immediate upon noting
that $\hat{U}=G$.
\qed

In view of Lemma~1, $\theta^*$ can
also be expressed as,
\be
\theta^*=\frac
{
    \pi\big(\hat{F}G-(P\hat{F})(PG)\big)
}
{
    E_\pi\Big[\Big(G(X_1)-PG(X_0)\Big)^2\Big]
}.
\label{eq:thetaStar3}
\ee

\subsection{A suboptimal empirical estimate of $\theta^*$}
\label{s:suboptimal}
Let $\{X_n\}$ be
a Markov chain
with transition kernel $P$ and invariant
measure $\pi$. In order to estimate the
mean $\pi(F)$ of some function $F$
under $\pi$, we replace the
ergodic averages $\mu_n(F)$ of (\ref{eq:ergodic})
by the modified estimates
$\mu_n(F_\theta)=\mu_n(F)-\theta\mu_n(U),$
where the control variate $U:=G-PG$
for some fixed function $G$, which,
we hope, approximates the solution $\hat{F}$
to the Poisson equation for $F$, or,
at least, is strongly correlated with $F$.
In order to select a ``good'' value for the
coefficient $\theta$ -- a value that
leads to a relatively small asymptotic variance
for the estimates $\mu_n(F_\theta)$ -- we
first 
consider the following simplistic scheme.

Pretending momentarily that $\{X_n\}$ is a sequence
of $\iid$ samples with distribution $\pi$, then
$\hat{F}=F$ and the optimal coefficient choice
for $\theta$ becomes,
$$\theta^*_{\tiid}
=\frac{\COV_\pi(F,G)}{\VAR_\pi(G)}
=\frac{\COV_\pi(F,U)}{\VAR_\pi(U)},$$
which can be adaptively estimated by,
$$\hat{\theta}_{n,\tiid}=
\frac{\mu_n(FU)}{\mu_n(U^2)}.$$
This leads us to the usual adaptive
estimator for $\pi(F)$, commonly used
in the case of $\iid$ samples,
\ben
\mu_{n,\tiid}(F)
:=
    \mu_n(F_{\hat{\theta}_{n,\tiid}})
=
    \mu_n\Big(F-
    \frac{\mu_n(FU)}{\mu_n(U^2)}U\Big)
=
    \mu_n(F)-
    \frac{\mu_n(U)\mu_n(FU)}{\mu_n(U^2)}.
\een
To examine its performance when used on
samples from a Markov chain,
we consider an example.

\medskip

\noindent
{\bf Example 1. A simple Gibbs sampler. }
Let $\pi(x,y)$ be
a bivariate Normal distribution
with zero mean, unit variances, and
covariance $\rho>0$.
We use the systematic-scan Gibbs sampler
to simulate from $\pi$. Starting from arbitrary
$X_0=x$ and $Y_0=y$,
$X_1$ is generated by sampling from 
$\pi(x|y)\sim N(\rho y,(1-\rho^2))$,
and then $Y_1$ is generated by sampling
from $\pi(y|X_1)\sim N(\rho X_1,(1-\rho^2))$.
Continuing this way produces
a Markov chain $\{(X_n,Y_n)\}$ with
distribution converging to $\pi$.

Suppose we wish to estimate the
expected value of $X^2$ under $\pi$.
Letting $F(x,y)=x^2$, the standard
estimates $\mu_n(F)\to\pi(F)=E_\pi(X^2)=1$
a.s., but when $\rho$ is close to 1 the
variance is high and the convergence
very slow. In this particularly simple
example, we can actually solve the
Poisson equation for $F$. Since $F$
is quadratic, we consider a candidate
solution of the form $G(x,y)=bx^2+cy^2$.
Direct calculation shows that,
$$PG(x,y)-G(x,y)=-bx^2+(\rho^4c +\rho^2b-c)y^2
    +(\rho^2c+b+c)(1-\rho^2),$$
and for this to be identically equal to
$-F(x,y)+\pi(F)=-x^2+1$, it suffices to take
$b=1$ and $c=\rho^2(1-\rho^4)^{-1}.$
Therefore,
$$\hat{F}(x,y)=x^2+\rho^2(1-\rho^4)^{-1}y^2.$$
From this we can compute the asymptotic variance
$\sigma_F^2$ of the estimates $\mu_n(F)$ by
substituting $\hat{F}$ in (\ref{eq:varF}),
to obtain,
$$\sigma_F^2=
E_\pi[(X^2+cY^2)^2-(1+cY^2)^2]
=2(1+\rho^4)(1-\rho^4)^{-1},$$
which is indeed high for $\rho\sim 1.$

And now suppose that,
as is typically the case in applications,
$\pi(x,y)$ is not available and we cannot
obtain an explicit solution for $\hat{F}$.
In order to create
a control variate $U$ it is natural to
start with $G=F$ itself, since we certainly
expect that $U=F-PF$ will be strongly
correlated with $F$. But $F$ only depends
on $x$, so, in order to take
advantage of the fact that we also
produce samples for $y$, we let
$G(x,y)=F(x,y)+F(y,x)=x^2+y^2$
and define the control variate,
$$U(x,y)=G(x,y)-PG(x,y)=x^2+(1-\rho^2-\rho^4)y^2
-(2-\rho^2-\rho^4).$$ We will now compare the performance of three
estimators: $(i)$~The standard estimator $\mu_n(F)$; $(ii)$~The
suboptimal adaptive estimator $\mu_{n,\tiid}(F)$ based on the
control variate $U$ defined above; and $(iii)$~The optimal
    estimator
    $\mu_{n}(F_{\theta^*})$
    based on the same control variate $U$,
    but with respect to the optimal
    value of $\theta^*$.

Since in this case we know both $\pi$ and
$\hat{F}$ explicitly, for the sake
of comparison we compute the theoretically
optimum value of $\theta$ appearing
in (\ref{eq:thetaStar}) as,
$$\theta^*=\frac{1+3\rho^2+2\rho^4}
{(1-\rho^4)(2+4\rho^2+3\rho^4+\rho^6)}.$$

\Figure{gibbsG} shows a typical realization
of the performance of all three
estimators for the following parameter values:
The correlation $\rho=0.95$, the number of steps
$n=5000$, the initial values are $x_0=y_0=0$,
and the optimal value of
$\theta^*\approx 3.273$.
In this experiment,
the (estimated) variance of the adaptive
estimator is smaller than that of the standard
estimator by a factor of $\approx 3.13$;
whereas the variance of the optimal estimator
is smaller than that of the standard estimator
by a factor of $\approx 18.52$.

\begin{figure}[ht!]
\centerline{\includegraphics[width=4.6in]{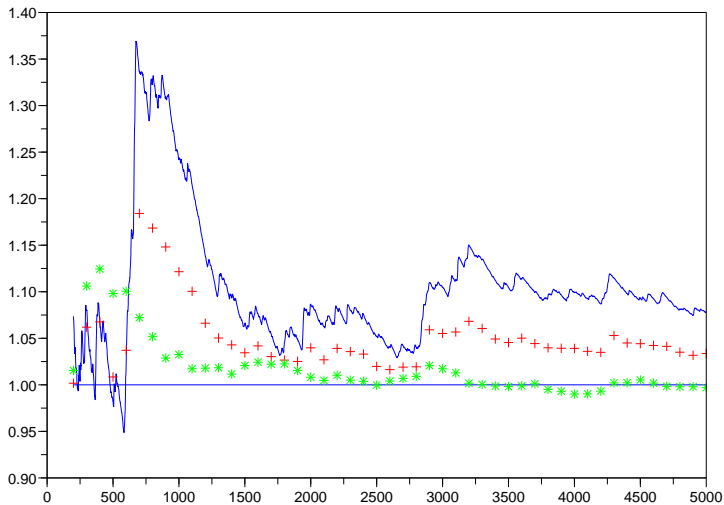}}
\caption{The sequence of the standard ergodic averages
is shown as a solid blue line; the suboptimal adaptive
estimates $\mu_{n,{\rm iid}}(F)$ as red ``+'' signs;
and the optimal estimates $\mu_n(F_{\theta^*})$
are green ``$*$'' signs. For visual clarity,
the estimates $\mu_{n,{\rm iid}}(F)$ and
$\mu_n(F_{\theta^*})$ are plotted only every 100 simulation
steps.}
\flabel{gibbsG}
\end{figure}

The reduction in the variance was computed from
$T=100$ independent repetitions of the same experiment.
For $\mu_n(F)$, we obtained $T=100$ different estimates
$\mu^{(i)}_n(F)$, for $i=1,2,\ldots,T$, and
the variance of $\mu_n(F)$ was estimated by,
\be
\frac{1}{T-1}\sum_{i=1}^{T}
[\mu^{(i)}_n(F)-\bar{\mu}_n(F)]^2,
\label{eq:red_est}
\ee
where
$\bar{\mu}_n(F)$ is the average of the $\mu_n^{(i)}(F)$.
The same procedure was applied to estimate the
variance of $\mu_{n,{\rm iid}}(F)$ and $\mu_n(F_{\theta^*})$.
The factors by which the variance of $\mu_n(F)$ is larger
than that of $\mu_{n,{\rm iid}}(F)$ and $\mu_n(F_{\theta^*})$,
respectively, are shown in Table~\ref{tab:ex1}.

\begin{table}[ht!]
  \begin{center}
    \begin{tabular}{|c||c|c|c|c|}
      \hline
      \multicolumn{5}{|c|}{\bf Variance reduction factors} \\
      \hline
      & \multicolumn{4}{|c|}{\em Simulation steps}\\
        \hline
    {\em Estimator} & $n=1000$ & $n=2000$ & $n=5000$ & $n=10000$\\
     \hline
$\mu_{n,{\rm iid}}(F)$ & 3.16 & 2.96 & 3.13 & 3.01 \\
     \hline
$\mu_n(F_{\theta^*})$  & 18.28 & 16.43 & 18.52 & 16.07 \\
     \hline
     \end{tabular}
  \end{center}
     \caption{Estimated factors by which the variance of $\mu_n(F)$
    is larger than that of $\mu_{n,{\rm iid}}(F)$ and
    $\mu_n(F_{\theta^*})$, respectively, after
    $n=1000,2000,5000$ and $10000$ simulation steps.}
     \label{tab:ex1}
\end{table}

For different values of the number of iterations $n$, the
corresponding variance reduction factors were computed based on
independent runs, and are not continuations of shorter runs. Note
that the adaptive estimates $\mu_{n,{\rm iid}}(F)$ were
actually computed in two steps: First the value for the coefficient
$\hat{\theta}_{n,\tiid}$ was computed, and then the values
$\mu_{n,{\rm iid}}(F)$ were calculated. In both passes, the same
simulation samples were used. We emphasize that 
{\em this procedure is
used throughout the paper}. 
Indeed, the fact that 
the estimators can be computed after the MCMC sample 
has been obtained is
a major advantage of our methodology.

Clearly, although the adaptive estimator $\mu_{n,\tiid}(F)$ does
offer a significant advantage over $\mu_n(F)$, there is a lot to be
gained from obtaining more accurate estimates of the optimal
coefficient $\theta^*$.  We remark that,
	instead of treating
    the samples $\{X_n\}$ as being $\iid$, more accurate
    estimates for $\theta^*$ can be obtained by approximating
    the expression (\ref{eq:thetaRatio}) via averages over
    blocks. Nevertheless, extensive simulation experiments
    clearly indicate that the corresponding estimation gains are
    usually negligible, while the optimal,
	consistent estimation procedures
	for $\theta^*$ 
	given in the following section make a very
    significant difference.

\subsection{Optimal empirical estimates of $\theta^*$ for
reversible chains}
\slabel{theta-rev}

Let $\Delta=P-I$ denote the generator of a
discrete time Markov chain $\{X_n\}$
with transition
kernel $P$.
If the chain is reversible, then $\Delta$
is a self-adjoint linear operator on the space
$L_2(\pi)$. This simply means that,
$$\pi(F\,\Delta G)=\pi(\Delta F\, G),$$
for any two functions $F,G\in L_2(\pi)$. Our central result in terms
of the estimation methodology is the observation that, in this case,
the optimal coefficient $\theta^*$ admits a representation that does
not involve the solution to Poisson's equation $\hat{F}$:

\medskip

\noindent
{\bf Proposition 1. } If the chain $\{X_n\}$ is
reversible, then the optimal coefficient
$\theta^*$ for the control variate
$U=G-PG$ can be expressed
as,
\be
\theta^*=
\theta^*_\trev:=\frac
{\pi\big((F-\pi(F))(G+PG)\big)}
{\pi(G^2-(PG)^2)},
\label{eq:theta_rev}
\ee
or, alternatively,
\be
\theta^*=
\theta^*_\trev:=\frac
{\pi\big((F-\pi(F))(G+PG)\big)}
    {E_\pi\Big[\Big(G(X_1)-PG(X_0)\Big)^2\Big]}.
\label{eq:theta_rev2}
\ee

\noindent
{\sc Proof. }
Let $\bar{F}=F-\pi(F)$ denote the centered
version of $F$, and recall that $\hat{F}$ solves
Poisson's equation for $F$, so $P\hat{F}=\hat{F}-\bar{F}$.
Therefore, the numerator in the expression
for $\theta^*$ in (\ref{eq:thetaStar})
can be expressed as,
\ben
\pi\big(\hat{F}G-(P\hat{F})(PG)\big)
&=&
\pi\big(\hat{F}G-(\hat{F}-\bar{F})(PG)\big)\\
&=&
\pi\big(\bar{F}PG-\hat{F}\Delta G\big)\\
&=&
\pi\big(\bar{F}PG-\Delta\hat{F} G\big)\\
&=&
\pi\big(\bar{F}PG+\bar{F} G\big)\\
&=&
\pi\big(\bar{F}(G+PG)\big).
\een
This proves (\ref{eq:theta_rev}),
and (\ref{eq:theta_rev2}) follows from
(\ref{eq:thetaStar3}).
\qed

The expressions
(\ref{eq:theta_rev})
and
(\ref{eq:theta_rev2})
immediately suggest estimating
$\theta^*$ via,
\ben
\hat{\theta}_{n,\trev,1}
&=&
    \frac
    {\mu_n(F(G+PG))-
    \mu_n(F)\mu_n(G+PG)}
    {\mu_n(G^2)-\mu_n((PG)^2)}\\
\mbox{or}
\;\;\;\;
\hat{\theta}_{n,\trev,2}
&=&
\frac
    {\mu_n(F(G+PG))-
    \mu_n(F)\mu_n(G+PG)}
{\frac{1}{n}\sum_{i=0}^{n-1}(G(X_i)-PG(X_{i-1}))^2}.
\een
The resulting estimators,
$\mu_n(F_{\hat{\theta}_{n,\trev,1}})$
and $\mu_n(F_{\hat{\theta}_{n,\trev,2}})$
for $\pi(F)$
based on the control variate $U=G-PG$
and the coefficients
$\hat{\theta}_{n,\trev,1}$
and
$\hat{\theta}_{n,\trev,2}$,
respectively,
are denoted,
\ben
\mu_{n,\trev,1}(F)
&:=&
    \mu_n(F_{\hat{\theta}_{n,\trev,1}})
    \;=\;\mu_n(F-\hat{\theta}_{n,\trev,1}U)\\
\mbox{and}
\;\;\;\;
\mu_{n,\trev,2}(F)
&:=&
    \mu_n(F_{\hat{\theta}_{n,\trev,2}})
    \;=\;\mu_n(F-\hat{\theta}_{n,\trev,2}U).
\een

\medskip

An alternative way for estimating $\theta^*$ adaptively, which also
applies to non-reversible chains, was recently developed in
\citet{meyn:book}, based on the ``temporal difference learning''
algorithm. As most of the chains we will consider are reversible and
this alternative method is computationally significantly more
expensive than our estimates $\hat{\theta}_{n,\trev,1}$ and
$\hat{\theta}_{n,\trev,2},$ it will not be
considered further in the present discussion. 

A slightly more general case of the earlier example with a bivariate
Gaussian density is considered below; the random-scan Gibbs sampler
is used to examine the performance of the two new estimators. We
note that although the systematic-scan Gibbs sampler in general does
not produce a reversible chain, the random-scan Gibbs sampler always
does. Also, the back-and-forth version of the systematic-scan Gibbs
sampler is reversible, see \citet{roberts-valencia4}.

\medskip

\noindent {\bf Example 2. The bivariate Gaussian through the
random-scan Gibbs sampler. } Let $(X,Y)\sim\pi(x,y)$ be an arbitrary
bivariate Normal distribution, where, without loss of generality, we
take the expected values of both $X$ and $Y$ to be zero and the
variance of $X$ to be equal to one. Let $\VAR(Y)=\tau^2$ and the
covariance $E(XY)=\rho\tau$ for some $\rho\in(-1,1)$. Given
arbitrary initial values $x_0=x$ and $y_0=y$, 
the Gibbs sampler selects one of the two
co-ordinates at random, and either
updates $y$ by sampling from $\pi(y|x)\sim N(\rho\tau
x,\,\tau^2(1-\rho^2))$, or $x$ 
from $N(\frac{\rho}{\tau}
y,\,1-\rho^2)$. Continuing this way produces a reversible Markov
chain $\{(X_n,Y_n)\}$ with distribution converging to $\pi$.

To estimate the
expected value of $X$ under $\pi$,
we let $F(x,y)=x$ and $G(x,y)=x+y$,
so that,
$$PG(x,y)=
\half(1+\rho\tau)x
+\half\Big(1+\frac{\rho}{\tau}\Big)y,$$
and the control variate $U=G-PG$ is,
$U(x,y)=G(x,y)-PG(x,y)=
\half(1-\rho\tau)x
+\half(1-\frac{\rho}{\tau})y.$
We will compare the performance
of five estimators:
$(i)$~The standard estimator $\mu_n(F)$;
$(ii)$~The suboptimal adaptive estimator $\mu_{n,\tiid}(F)$
    based on the control variate $U=G-PG$ defined
    above;
$(iii,iv)$~The two adaptive estimators
    $\mu_{n,\trev,1}(F)$
    and $\mu_{n,\trev,2}(F)$
    for the same control variate $U$;
$(v)$~The {\em optimal} estimator
    $\mu_{n}(F_{\theta^*})$
    based on the same control variate,
    but with respect to the optimal
    value of $\theta^*$.

In Figure~\ref{f:example2a} we plot the
results of all five estimators,
applied to
a typical execution of the Gibbs
sampler with $n=5000$ steps
and initial values $x_0=y_0=0.1$.
The problem parameter
values are $\rho=0.99$ and
$\tau^2=10$.

\begin{figure}[ht!]
\centerline{\includegraphics[width=4.6in]{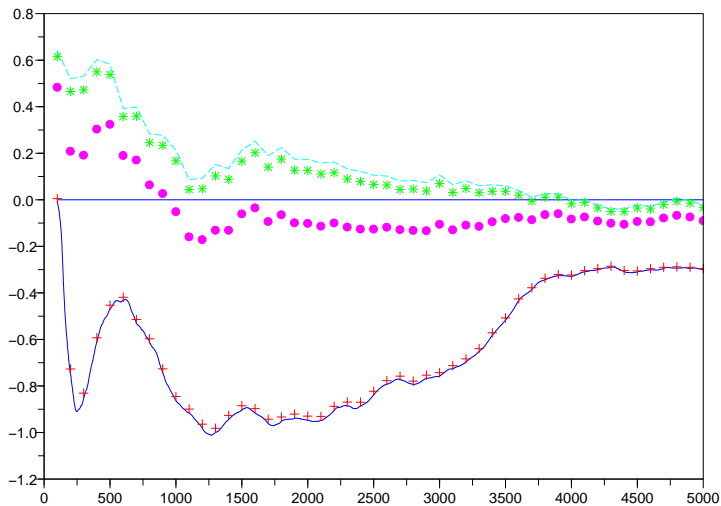}}
\caption{The sequence of the standard ergodic averages
$\mu_n(F)$
is shown as a solid blue line; the suboptimal adaptive
estimates $\mu_{n,{\rm iid}}(F)$ as red ``+'' signs;
the optimal adaptive estimates
$\mu_{n,\trev,1}(F)$ as bold magenta dots,
$\mu_{n,\trev,2}(F)$ as a dashed cyan line,
and the estimates $\mu_n(F_{\theta^*})$
corresponding to the optimal value of $\theta^*$
as green ``$*$'' signs. For visual clarity,
all estimates except $\mu_{n}(F)$
are plotted only every 100 simulation
steps.}
\flabel{example2a}
\end{figure}

While the optimal estimator
$\mu_n(F_{\theta^*})$
offers an obviously large advantage
over the standard estimates $\mu_n(F)$,
the improvement of the suboptimal estimator
$\mu_{n,{\rm iid}}(F)$ is rather insignificant.
The adaptive estimators
$\mu_{n,\trev,1}(F)$ and
$\mu_{n,\trev,2}(F)$ are similarly very effective,
and their performance is
fairly close to that of the optimal $\mu_n(F_{\theta^*}).$
As in Example~1, we compute the factor by which
each of these methods reduces the variance of the
standard estimates $\mu_n(F)$, using
$T=200$ independent repetitions of the same experiment;
recall equation (\ref{eq:red_est}) above.
The results are shown in Table~\ref{tab:ex2}.

\begin{table}[ht!]
  \begin{center}
    \begin{tabular}{|c||c|c|c|c|c|}
      \hline
      \multicolumn{6}{|c|}{\bf Variance reduction factors} \\
      \hline
      & \multicolumn{5}{|c|}{\em Simulation steps}\\
        \hline
    {\em Estimator} & $n=1000$& $n=5000$& $n=10000$& $n=50000$& $n=100000$\\
     \hline
$\mu_{n,{\rm iid}}(F)$  & 1.04 & 1.03 & 1.02 & 1.02 & 1.02\\
     \hline
$\mu_{n,\trev,1}(F)$    & 2.14 & 6.25 & 6.77 & 8.26 & 7.50\\
     \hline
$\mu_{n,\trev,2}(F)$    & 2.79 & 5.66 & 6.58 & 8.19 & 7.54\\
     \hline
$\mu_n(F_{\theta^*})$   & 5.23 & 9.12 & 8.20 & 8.25 & 7.53\\
     \hline
     \end{tabular}
  \end{center}
     \caption{Estimated factors by which the variance of $\mu_n(F)$
    is larger than that of
    $\mu_{n,{\rm iid}}(F)$, $\mu_{n,\trev,1}(F)$
    $\mu_{n,\trev,2}(F)$ and
    $\mu_n(F_{\theta^*})$, respectively, after
    $n=1000,5000,10000,50000$ and $100000$ simulation steps.}
     \label{tab:ex2}
\end{table}

Clearly, both adaptive estimators
$\mu_{n,\trev,1}(F)$, $\mu_{n,\trev,2}(F)$
perform very well, and their results are reasonably
close to those of the optimal estimator $\mu_n(F_{\theta^*})$.
A natural way to attempt to obtain an even greater
improvement in terms of variance reduction
would be to consider a control variate $U$ based
on a $G$ of the form $G(x,y)=ax+by$,
for coefficients $a\neq b$. But there is no
obvious choice for the relationship between these
two coefficients, and also we do not want to develop
methods that are too model-specific. A generic way
to address such problems is to consider two control
variates $U_1$, $U_2$, based on the two different
functions $G_1(x,y)=x$ and $G_2(x,y)=y$.
The corresponding methodology for such cases
is developed in Section~\ref{s:multiple}, where
we also revisit this example.

Another well-known difficulty with the standard estimates in this
example (in addition to their high variance) is that they converge
very slowly when the initial values of the Gibbs sampler are far
from their mean. The above results are from simulations with
$x_0=y_0=0.1$, and we also run several examples with different
initial values. In those cases we found that a lot of the time the
estimators $\mu_{n,\trev,1}(F)$ and $\mu_{n,\trev,2}(F)$ not only
reduced the variance, but also greatly improved the bias. An example
with initial values $x_0=4$ and $y_0=12$ is shown in
Figure~\ref{f:example2c}. A more detailed discussion of this issue
will be given in Section~\ref{s:bias}, where we also address the
question of choosing between the two adaptive estimators,
$\mu_{n,\trev,1}(F)$ and $\mu_{n,\trev,2}(F)$.

\begin{figure}[ht!]
\centerline{\includegraphics[width=4.6in]{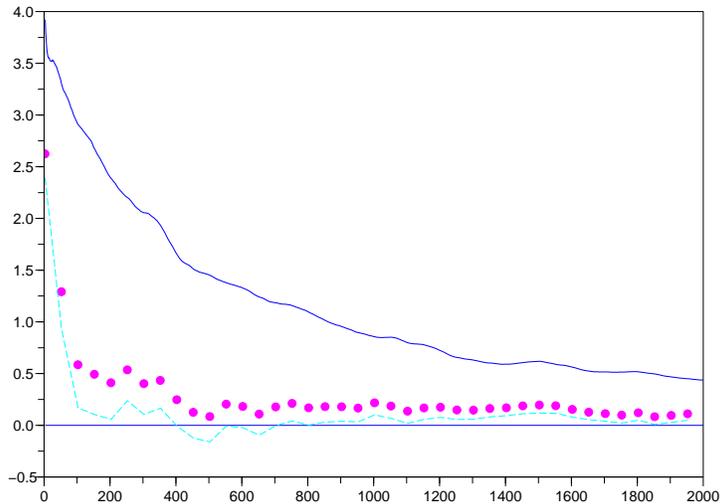}}
\caption{The sequence of the standard ergodic averages
is shown as a solid blue line;
the adaptive estimates
$\mu_{n,\trev,1}(F)$,
plotted only every 50 simulation steps,
are shown as bold magenta dots,
and the adaptive estimates
$\mu_{n,\trev,2}(F)$ as a dashed
cyan line.
}
\flabel{example2c}
\end{figure}


\section{Five Simple MCMC Examples}
\label{sec:simple}

Below we present five examples more closely motivated by problems in
statistical inference.  Among the vast array of simple MCMC
samplers that can be used for illustration purposes, 
we have chosen a set of representative examples that
cover a broad class of real applications.  The Gaussian-Gamma 
posterior in Example~3, as well as the 
the bivariate Gaussian density of Example~2,
are representative of the large class of normal hierarchical 
models that are analyzed in \citet{hills}. Similarly, the Gibbs 
sampler of Example~4 is seen as a simplistic version of a wide 
class of models that include discrete variables as latent variable 
indicators or model indices. The discrete state space random-walk 
Metropolis algorithm in Example~5 is used in model search algorithms in 
which analytical or approximate integration of all model parameters 
is first performed;
see for example \citet{Clyde04}.  A simple version 
of a finite-mixtures mode of Normals is explored 
in detail in Example~6. This
this class of models has been, and still is, 
one of the most challenging inference problems.  
Finally, we illustrate our methodology in the case
of a ``difficult'' model where Cauchy priors result in
heavy-tailed posterior densities; such densities 
are commonly met in, for example, spatial statistics; see 
\citet{DelRob03} for an illustrative example.

\medskip

\noindent
{\bf Example 3. A Gaussian-Gamma posterior. }
First we consider an example of the random-scan Gibbs
sampler applied to simple Bayesian inference problem.
The model is a simple two-parameter example
of \citet{gilks:96},
in which we begin with observations
$x=(x_1,x_2,\ldots,x_N)$
that are independently generated from
a $N(\mu,\gamma^{-1})$ distribution,
and we place priors
$\mu\sim N(0,1)$ and $\gamma\sim \text{Gamma}(2,1)$
on the parameters $\mu$ and $\gamma$,
respectively.
[Throughout the paper, the paramatrization of the
Gamma$(a,b)$ density is chosen so that it has mean $a/b$.]
It is straightforward that
Gibbs sampling from the posterior $\pi(\mu,\gamma|x)$
proceeds by updating $\mu$ given $\gamma$
from a Normal density with mean $(\gamma\sum_i x_i)/(1+N\gamma)$ and
variance $1/(1+N\gamma)$, and $\gamma$ given $\mu$
from a Gamma density with index $2+N/2$ and
scale $1+\half\sum_i(x_i-\mu)^2$.
In our simulation, we assume that $N=10$ and that
the data vector $x=(x_i)$
is given by $x=(-23,27,12,17,-8,2,-18,17,7,-33)$, so that
the sample mean is zero.
We wish to estimate the posterior mean
of $\mu$, so we let $F(\mu,\gamma)=\mu$.
Although in general this posterior mean
is not computable in closed form, here
the posterior marginal density of $\mu$
is proportional to the product of a Student's $t$
density with mean zero (because the
sample mean of $x$ is zero)
and the prior $N(0,1)$ density.
Therefore, the resulting
density is symmetric around zero, which implies
that the posterior mean of $\mu$ is actually zero.
We compare the performance of the
simple empirical averages $\mu_n(F)$ with
the adaptive estimators
$\mu_{n,\trev,1}(F)$
and $\mu_{n,\trev,2}(F)$,
based on the control variate $U=G-PG$ with
$G(\mu,\gamma)=\mu$.

\begin{figure}[ht!]
\centerline{\includegraphics[width=4.6in]{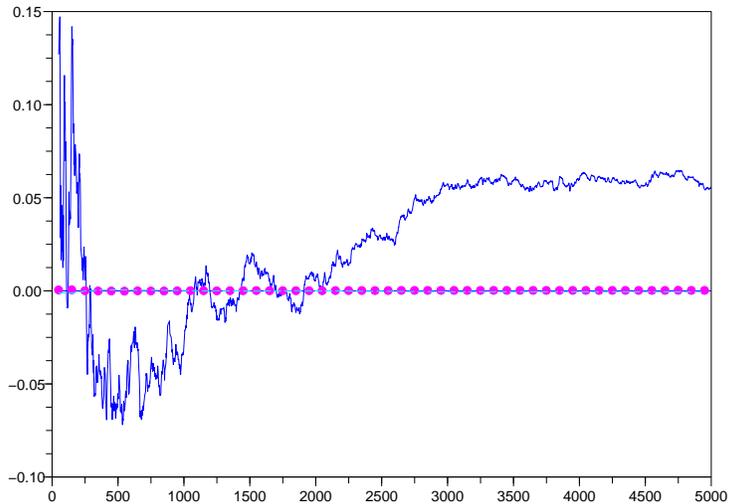}}
\caption{The sequence of the standard ergodic averages
is shown as a solid blue line;
the adaptive estimates
$\mu_{n,\trev,1}(F)$ as bold magenta dots;
and the adaptive estimates
$\mu_{n,\trev,2}(F)$ as a cyan dashed line.
For visual clarity,
the values $\mu_{n,\trev,1}(F)$
are plotted only every 100 simulation
steps.}
\flabel{example3}
\end{figure}

Figure~\ref{f:example3} shows a typical
random-scan Gibbs sampling run of length $n=5000$,
with starting values $\mu_0=\gamma_0=1$.
It is obvious from the plot that both
adaptive estimators converge incredibly
fast compared to the standard ergodic
averages $\mu_n(F)$. The corresponding variance
reduction factors, computed from $T=100$ repetitions
of the same experiment, are shown in Table~\ref{tab:ex3}.
\begin{table}[ht!]
  \begin{center}
    \begin{tabular}{|c||c|c|c|c|c|}
      \hline
      \multicolumn{5}{|c|}{\bf Variance reduction factors} \\
      \hline
      & \multicolumn{4}{|c|}{\em Simulation steps}\\
        \hline
{\em Estimator}&$n=1000$&$n=5000$&$n=10000$& $n=50000$\\
     \hline
$\mu_{n,\trev,1}(F)$  & 9403 & 341095 & 419766 & 20453186 \\
     \hline
$\mu_{n,\trev,2}(F)$  & 713 &  1880 & 5287    & 15495 \\
     \hline
     \end{tabular}
  \end{center}
     \caption{Estimated factors by which the variance of $\mu_n(F)$
    is larger than that of
    $\mu_{n,\trev,1}(F)$
    and $\mu_{n,\trev,2}(F)$,
    after
    $n=1000,5000,10000,$ and $50000$ simulation steps.}
     \label{tab:ex3}
\end{table}

Given the tremendous effectiveness of the control variate $U=G-PG$
with $G(\mu,\gamma)=\mu$, it is natural to ask if perhaps a multiple
of $G$ actually solves the Poisson equation, that is, if
$\theta^*(G-PG)=F-\pi(F)$. Since $\pi(F)=0$ and in the simulation
experiments both $\hat{\theta}_{n,\trev,1}$ and
$\hat{\theta}_{n,\trev,2}$ apparently converge to values very close
to 2, we examine the relationship, $2(G-PG)=\mu$. Substituting the
expressions for $G$ and $PG$ this becomes, \be \frac{\gamma\sum_i
x_i}{1+N\gamma} \eqquestion 0, \label{eq:QPoisson} \ee which, in our
case, is indeed an equality, since the empirical mean of our sample
$x$ is equal to zero. More generally, this will be an approximate
equality (at least for most of the relevant values of $\mu$ and
$\gamma$), as long the empirical mean of the sample is close to
zero, or if most of the mass of the posterior on $\gamma$ is
concentrated near zero. In fact, a multiple of $G(\mu,\gamma)=\mu$
will always solve the Poisson equation exactly, as long as the mean
of the Gaussian prior on $\mu$ instead of zero is taken to be equal
to the sample mean of $x$.

The above discussion explains the effectiveness of the control
variate $U=G-PG$ with $G(\mu,\gamma)=\mu$, but it also suggests that
if the number of observations $N$ is small, and either: $(a)$~the
sample mean of the observations $x$ is not close to zero; or~$(b)$
the empirical standard deviation of $x$ is not appropriately
``small'' (in other words, the posterior on $\gamma$ is not
concentrated near zero); then this $G$ would  not be an approximate
solution to the Poisson equation, and the corresponding control
variate $U$ would be much less effective. Nevertheless, even in the
unlikely scenario where the observation vector is $x'=(x'_i)= (4.75,
5.09, 4.63, 4.73, 5.08, 4.47, 5.24, 5.06, 4.98, 5.21)$, using the
same control variate as before is quite effective; see the
corresponding results in Table~\ref{tab:ex3b}.

\begin{table}[ht!]
  \begin{center}
    \begin{tabular}{|c||c|c|c|c|c|c|c|}
      \hline
      \multicolumn{7}{|c|}{\bf Variance reduction factors} \\
      \hline
      & \multicolumn{6}{|c|}{\em Simulation steps}\\
        \hline
{\em Estimator}&$n=1000$&$n=5000$&$n=10000$& $n=50000$& $n=100000$& $n=200000$\\
     \hline
$\mu_{n,\trev,1}(F)$  & 0.02 & 0.01 & 0.00 & 0.44 & 3.43 & 3.87 \\
     \hline
$\mu_{n,\trev,2}(F)$  & 0.37 & 4.46 & 4.17 & 8.81 & 7.88 & 6.50 \\
     \hline
     \end{tabular}
  \end{center}
     \caption{Estimated factors by which the variance
    of $\mu_n(F)$ is larger than that of
    $\mu_{n,\trev,1}(F)$
    and $\mu_{n,\trev,2}(F)$,
    after
    $n=1000,5000,10000,50000,100000$ and $200000$ simulation steps.
    Here the vector $x$ of observations has sample mean 4.924
    and sample standard deviation $\approx 0.068$.}
     \label{tab:ex3b}
\end{table}

The reason this scenario is referred to as
being ``unlikely''
is because the set of observations $x'$ was
actually obtained as an i.i.d.\ sample
(rounded off to two decimal places) from
the $N(5,0.09)$ density; it has sample mean equal to 4.924,
and sample variance $\approx 0.068$.
Therefore, having a $N(0,1)$ prior on the mean
of the observations $x'$ that actually vary
between 4.47 and 5.24 is an unreasonable
choice. Also note that
both of the potential sources of concern~$(a)$, $(b)$ above
apply here. Indeed, for $\gamma=1/0.068$,
the right-hand-side of (\ref{eq:QPoisson}) is
$\approx 4.9$, which is certainly not close to zero.
Still, using the control variate $U=G-PG$ with
$G(\mu,\gamma)=\mu$ consistently yields nontrivial
variance reduction factors.

\medskip


\noindent {\bf Example 4. An example with a discrete variable. }
Next we construct a bivariate density with a discrete variable $z$
and a continuous variable $p$, where $z|p \sim {\rm Bern}(p)$ and
$p\sim {\rm Beta}(\alpha,\beta)$. The random-scan Gibbs sampler
draws randomly from either $z|p\sim{\rm Bern}(p)$ or from $p|z \sim
{\rm Beta}(\alpha+z,\beta+1-z)$.

We wish to estimate the mean of $z$, so we set $F(z,p)=z$ and
examine the performance of the ergodic averages $\mu_n(F)$ and the
two adaptive estimators $\mu_{n,\trev,1}(F)$ and
$\mu_{n,\trev,2}(F)$ based on the control variate $U=G-PG$; for $G$
we take, as before, $G(z,p)=z+p$. Figure~\ref{f:example4} depicts a
typical realization of the random-scan Gibbs sampler, with
$\alpha=2$, $\beta=1$, starting values $z_0=p_0=1/2$, and $n=5000$
steps. Here, the true value of $\pi(z)$ is
$\alpha/(\alpha+\beta)=2/3$.
The corresponding variance reduction factors,
estimated from $T=100$ repetitions of the same experiment,
are shown in Table~\ref{tab:ex4}.

\begin{figure}[ht!]
\centerline{\includegraphics[width=4.6in]{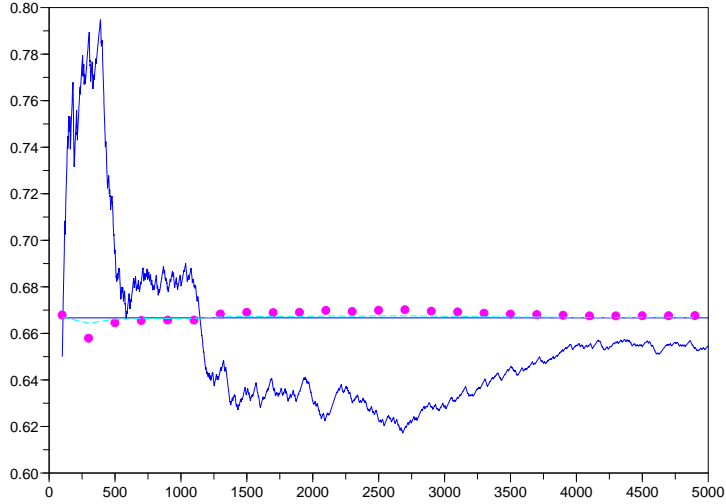}}
\caption{The sequence of the standard ergodic averages
is shown as a solid blue line;
the adaptive estimates
$\mu_{n,\trev,1}(F)$ as bold magenta dots;
and the adaptive estimates
$\mu_{n,\trev,2}(F)$ as a cyan dashed line.
For visual clarity,
the values $\mu_{n,\trev,1}(F)$
are plotted only every 200 simulation
steps.}
\flabel{example4}
\end{figure}

\begin{table}[ht!]
  \begin{center}
    \begin{tabular}{|c||c|c|c|c|c|c|}
      \hline
      \multicolumn{7}{|c|}{\bf Variance reduction factors} \\
      \hline
      & \multicolumn{6}{|c|}{\em Simulation steps}\\
        \hline
{\em Estimator}&$n=1000$&$n=5000$&$n=10000$& $n=20000$& $n=50000$& $n=100000$\\
     \hline
$\mu_{n,\trev,1}(F)$    & 5.89  & 24.50  & 41.40  & 212.8  & 702.5 & 1721.3\\
     \hline
$\mu_{n,\trev,2}(F)$    & 247.4 & 1286.5 & 2145.8 & 4235.4 & 12066 & 24777 \\
     \hline
     \end{tabular}
  \end{center}
     \caption{Estimated factors by which the variance of $\mu_n(F)$
    is larger than that of
    $\mu_{n,\trev,1}(F)$
    and $\mu_{n,\trev,2}(F)$,
    after
    $n=1000,5000,10000,20000,50000$ and $100000$ simulation steps.}
     \label{tab:ex4}
\end{table}


Like in Example~3, since the use of the control
variate $U$ decreases the MCMC variance dramatically,
it is natural to check if perhaps a multiple of $G$
solves the Poisson equation. Direct
computation gives,
$$PG(z,p)=
p
+
\frac{\alpha+(\alpha+\beta+2)z}{2(\alpha+\beta+1)},
$$
so that,
$$PG(z,p)-G(z,p)=
\frac{\alpha+\beta}{2(\alpha+\beta+1)}
\Big[
-z+\frac{\alpha}{\alpha+\beta}
\Big]
=
\frac{\alpha+\beta}{2(\alpha+\beta+1)}
\Big[-F(z,p)+\pi(F)
\Big].
$$
Indeed, then, $G$ is a multiple of the
solution of the Poisson equation for $F$,
$$
\hat{F}(z,p)
=
\frac
{2(\alpha+\beta+1)}{\alpha+\beta}(z+p)
=
\frac{2(\alpha+\beta+1)}{\alpha+\beta}\,G(z,p),
$$
and the optimal coefficient for $U$ is,
$$
\theta^* =
\frac
{2(\alpha+\beta+1)}{\alpha+\beta}.
$$
This explains the effectiveness of this particular choice of the
function $G$. Incidently, it is somewhat remarkable that a multiple
of the same function $G(z,p)=z+p$ solves the Poisson equation for
any choice of the parameter values $\alpha,\beta$.

\medskip

\noindent
{\bf Example 5. Random-walk Metropolis for Poisson generation. }
Consider the target distribution
$\pi\sim\mbox{Poisson}(\lambda)$.
When the mean $\lambda$ is large, it is hard
to sample from $\pi$ directly and,
instead, we consider a random-walk Metropolis
sampler, which, given $X_n=x$, proposes a move
to $X'_{n+1}=x+Z_n$, where the increments $Z_n$
are i.i.d.\ and
$Z_n=-1$ or $+1$ with probability $1/2$ each.
The acceptance probability can be easily computed
as,
$$\alpha(x,y)
=
\begin{cases}
\min\{1, \frac{\lambda}{x+1}\}  & \text{if $y=x+1$,}\\
\min\{1, \frac{x}{\lambda}\}    & \text{if $y=x-1$.}
\end{cases}
$$
Suppose we wish to estimate the mean of
$\sqrt{x}$ under $\pi$, so let $F(x)=\sqrt{x}$.
To use a control variate $U=G-PG$ with respect
to some function $G$ on the integers, note
that $PG$ is,
$$
PG(x)=
G(x)
+\frac{1}{2}\alpha(x,x+1)[G(x+1)-G(x)]
+\frac{1}{2}\alpha(x,x-1)[G(x-1)-G(x)],
$$
so that, in particular,
taking $G(x)=x$, we have,
$$U(x) = \half\alpha(x,x-1)-\half\alpha(x,x+1).$$

Figure~\ref{f:MHPoisson} shows a typical realization of the
Metropolis sampler, using $G(x)=x$, with initial value $x_0=95$, for
$n=10000$ simulation steps. The ``true'' mean of $\sqrt{X}$ under
$\pi$ is estimated to be $\approx 9.9874$, after 3 million
Metropolis steps.
The corresponding variance reduction factors,
estimated from $T=100$ repetitions of the same experiment, are shown
in Table~\ref{tab:MHPoisson}.

\begin{figure}[ht!]
\centerline{\includegraphics[width=4.4in]{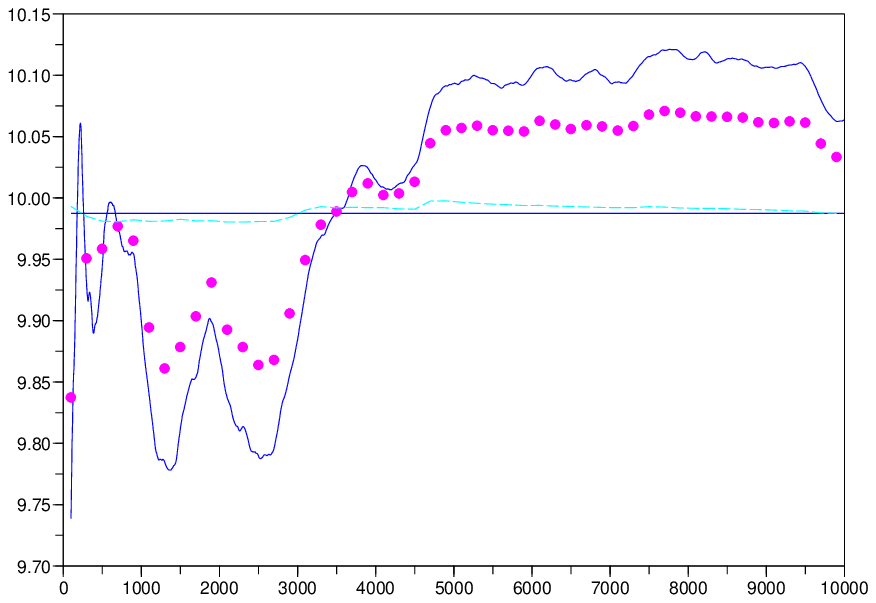}}
\caption{The sequence of the standard ergodic averages
is shown as a solid blue line and
the adaptive estimates
$\mu_{n,\trev,2}(F)$
as a dashed cyan line.
The adaptive estimates
are plotted only every 200 simulation
steps.}
\flabel{MHPoisson}
\end{figure}

\begin{table}[ht!]
  \begin{center}
    \begin{tabular}{|c||c|c|c|c|}
      \hline
      \multicolumn{5}{|c|}{{\bf Variance reduction factors}}\\
      \hline
      & \multicolumn{4}{|c|}{\em Simulation steps}\\
        \hline
{\em Estimator}& $n=1000$& $n=10000$& $n=50000$& $n=100000$\\
     \hline
$\mu_{n,\trev,1}(F)$    & 0.91 & 0.32  & 0.15  & 0.20 \\
     \hline
$\mu_{n,\trev,2}(F)$    & 4.73 & 39.19 & 157.5 & 239.98 \\
     \hline
     \end{tabular}
  \end{center}
     \caption{Estimated factors by which the
    variance of $\mu_n(F)$
    is larger than the corresponding
    variance of
    $\mu_{n,\trev,1}(F)$ and
    $\mu_{n,\trev,2}(F)$,
    respectively,
    after
    $n=1000,10000, 50000$ and $100000$ simulation
    steps.}
     \label{tab:MHPoisson}
\end{table}



\noindent {\bf Example 6. A two-parameter Gaussian mixture posterior. }
We examine a simple Gaussian mixture example as in
\citet[Example~9.2]{robert:book}. Suppose
$x=(x_1,x_2,\ldots,x_N)$
are independent observations
from the mixture 
$pN(\mu_1,\sigma^2)+(1-p)N(\mu_2,\sigma^2)$;
the mixing proportion $p$ and the variance
$\sigma^2$ are assumed fixed and known,
and $N(0,10\sigma^2)$ priors
are placed on the means $\mu_1,\mu_2$.
To facilitate sampling from the posterior,
the model can alternatively
be described in terms of latent variables
$Z=(Z_1,Z_2,\ldots,Z_N)$, where the $Z_i$ are
independent with distribution $P(Z_i=1)=1-P(Z_i=0)=p,$
and, conditional on $\mu_1,\mu_2$ and $Z$, each
$X_i|Z_i=1\sim N(\mu_1,\sigma^2)$, and
$X_i|Z_i=0\sim N(\mu_2,\sigma^2)$.

Conditional on $x$ and $z$, the parameters
$\mu_1$ and $\mu_2$ are independent,
with,
\be
\pi(\mu_1|x,z)
&\sim&
    N\left(\frac{\sum_{j}z_jx_j}{n_1+1/10},
    \frac{\sigma^2}{n_1+1/10}\right),
    \nonumber\\
\pi(\mu_2|x,z)
&\sim&
    N\left(\frac{\sum_{j}(1-z_j)x_j}{n_2+1/10},
    \frac{\sigma^2}{n_2+1/10}\right),
    \label{eq:mu_full}
\ee
respectively,
where
$n_1=\sum_jz_j$ is the number of $z_i$ that are equal
to 1, and $n_2=\sum_j(1-z_j)=N-n_1$ is the number of
$z_i$ that are equal to zero.
Also, given $\mu_1,\mu_2$ and $x$,
the $Z_i$ are independent, and
for each $i=1,2,\ldots,N$,
\be
\pi(Z_i=1|x,\mu_1,\mu_2)
=q_i:=
    \frac
    {
    p\exp\{-(x_i-\mu_1)^2/2\sigma^2\}
    }
    {
    p\exp\{-(x_i-\mu_1)^2/2\sigma^2\}
    +(1-p)\exp\{-(x_i-\mu_2)^2/2\sigma^2\}
    }.
\;\;\;
\label{eq:qi}
\ee
The random-scan Gibbs sampler here draws
a sample from $\mu_1$, $\mu_2$, or from the
entire vector $Z$, each chosen with probability $1/3$.


We consider an example with the exact same parameter settings as in
\citet[Example~9.2]{robert:book}: With $p=0.7$, $\sigma^2=1$,
$\mu_1=0$ and $\mu_2=2.7$, we generated $N=500$ samples from the
mixture $pN(\mu_1,\sigma^2)+(1-p)N(\mu_2,\sigma^2)$. In order to
estimate $\mu_1$, the function $F$ is set to be
$F(\mu_1,\mu_2,Z)=\mu_1$.
Using a control variate $U=G-PG$ with the simplest
choice of $G=F$, yields variance reduction
factors 
around 4, which are significantly 
smaller than those achieved in some of the earlier examples.
For that reason, we also consider 
$G(\mu_1,\mu_2,Z)=\sum_iZ_ix_i$ as a
different candidate $G$ for the control
variate $U$. In fact, we let
$G=c\mu+\sum_iZ_ix_i$, and select the value of $c$
so that a multiple of $G$ is as close as possible
to a solution of the Poisson equation, that is,
$\theta(PG-G)\approx -F+\pi(F)$, for some $\theta$;
substituting the values of $F$, $G$ and $PG$,
and taking $\pi(F)$ to be equal to the prior
expectation of $F$ (namely, zero),
this becomes,
\be
\sum_ix_i\Big[Z_i\Big(1-\frac{c}{n_1+1/10}\Big)
-q_i \Big]
\approx \Big(\frac{3}{\theta}-c\Big)\mu_1,
\label{eq:Ptry}
\ee
where the $(q_i)$ are the Bernoulli parameters
of the $(Z_i)$, given in (\ref{eq:qi}).
A reasonable goal here is to choose $c$ so as
to reduce the variability of the left-hand-side
as much as possible, since it is not directly related
to $\mu_1$. Ideally,
this would mean taking $c=n_1+1/10$, but since
$n_1$ is itself random, we take $c$ to be equal to the
(prior) expectation of that expression, namely,
$c=(Np+1/10)$, so that the resulting function $G$ is,
$$G(\mu_1,\mu_2,Z)=(Np+1/10)\mu_1+\sum_iZ_ix_i.$$

A typical realization of the
estimates based on $n=10000$ Gibbs steps is shown in
Figure~\ref{f:example5b}, and the corresponding
variance reduction factors are displayed in
Table~\ref{tab:ex5i}.
The initial values are $\mu_1=0$,
$\mu_2=1$, and the ``true'' posterior mean 
of $\mu_1$ is estimated to be $\approx -0.0143$,
after 10 million Gibbs steps.

\begin{figure}[ht!]
\centerline{\includegraphics[width=4.6in]{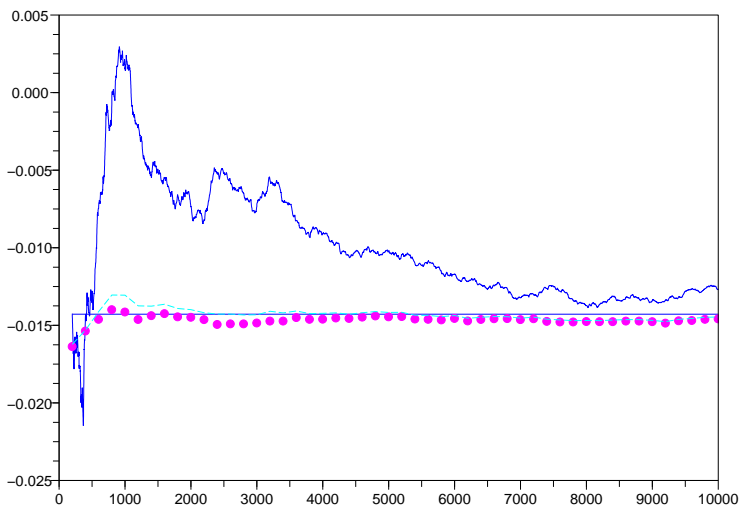}}
\caption{The sequence of the standard ergodic averages
is shown as a solid blue line;
the adaptive estimates
$\mu_{n,\trev,1}(F)$ as bold magenta dots;
and the adaptive estimates
$\mu_{n,\trev,2}(F)$ as a cyan dashed line.
For visual clarity,
the values $\mu_{n,\trev,1}(F)$
are plotted only every 200 simulation
steps.}
\flabel{example5b}
\end{figure}

\begin{table}[ht!]
  \begin{center}
    \begin{tabular}{|c||c|c|c|c|}
      \hline
      \multicolumn{5}{|c|}{\bf Variance reduction factors} \\
      \hline
      & \multicolumn{4}{|c|}{\em Simulation steps}\\
        \hline
{\em Estimator}& $n=10000$& $n=50000$& $n=100000$& $n=200000$\\
     \hline
$\mu_{n,\trev,1}(F)$ & 11.76 & 15.81 & 19.02 & 22.12 \\
     \hline
$\mu_{n,\trev,2}(F)$ & 11.63 & 15.44 & 18.98 & 21.98 \\
     \hline
     \end{tabular}
  \end{center}
     \caption{Estimated factors by which the
    variance of $\mu_n(F)$
    is larger than the corresponding
    variances of
    $\mu_{n,\trev,1}(F)$
    and $\mu_{n,\trev,2}(F)$,
    respectively,
    after
    $n=10000,50000,100000$ and $200000$ simulation steps.}
     \label{tab:ex5i}
\end{table}

Incidently, the above calculation
suggests that the optimal value for $\theta$
here would be the one that also makes the right-hand-side
of (\ref{eq:Ptry}) vanish, namely,
$\theta^*\approx 3/c=3/(Np+1/10)\approx 0.009$.
In our simulation experiments, the estimates
of $\theta^*$ produced by both $\hat{\theta}_{n,\trev,1}$
and $\hat{\theta}_{n,\trev,2}$ are around
$0.011$, which is indeed quite close.

Finally we note that models of this type often present a
difficultly, in that the posterior on $(\mu_1,\mu_2)$ is bimodal. As
a result, if the Gibbs sampler is initialized near the lower mode,
it will never visit the neighborhood of the actual mode, at least
not in any reasonable amount of time; see
\citet{robert:book,diebolt-robert:94}. A more general Gaussian
mixture model that at least partially addresses this issue is
explored in Section~\ref{s:Gcomplex}.


\medskip

\noindent
{\bf Example 7. A Metropolis-within-Gibbs sampler. }
We consider an inference problem motivated by
a simplified version of an example
in \citet{roberts-rosenthal:pre}.
Suppose $N$ i.i.d.\ observations
$x=(x_1,x_2,\ldots,x_N)$ are drawn from a $N(\phi  ,V)$
distribution, and place independent priors
$\phi  \sim\mbox{Cauchy}(0,1)$ and $V\sim\mbox{IG}(1,1)$,
on the parameters $\phi  ,V$, respectively.
The induced full conditionals of the posterior are
easily seen to satisfy,
\ben
\pi(\phi  |V,x)
&\propto&
    \Big(\frac{1}{1+\phi  ^2}\Big)\exp\Big\{
    -\frac{1}{2V}\sum_i(\phi  -x_i)^2
    \Big\},\\
\mbox{and}\;\;\;\;
\pi(V|\phi  ,x)
&\sim&
    \mbox{IG}
    \Big(1+\frac{N}{2},1+\frac{1}{2}\sum_i(\phi  -x_i)^2\Big).
\een Since the distribution $\pi(\phi  |V,x)$ is not of standard
form, direct Gibbs sampling is not possible. Instead, we use a
random-scan Metropolis-within-Gibbs sampler,
\citet{muller:93,tierney:94}, and either update $V$
from its conditional (Gibbs step), or update $\phi  $ in a random
walk-Metropolis step with a $\phi  '\sim N(\phi  ,1)$ proposal, each
case chosen with probability $1/2$.
Since both the Cauchy and the inverse Gamma distributions are heavy
tailed, we naturally expect that the MCMC samples will be highly
variable. Indeed, this was found to be the case in the simulation
example we consider next, where the above algorithm is applied to a
vector $x$ of $N=100$ i.i.d.\ $N(2,4)$ observations, and with
initial values $\phi_0=0$ and $V_0=1$. 
As a result of this
variability, the standard empirical averages of the values of the
two parameters also converge very slowly. Since $V$ is the more
variable of the two, we let $F(\phi  ,V)=V$ and consider the problem
of estimating its posterior mean. We will compare the performance of
the standard empirical averages $\mu_n(F)$ with that of the two
adaptive estimators $\mu_{n,\trev,1}(F)$ and $\mu_{n,\trev,2}(F)$,
with the control variate $U=G-PG$ defined in terms of the function
$G=F$. Note that, in this setting, we are restricted in our choice
of functions $G$ to those which depend only on $V$. Since $\phi $ is
updated by an accept/reject Metropolis step, if $G$ depended on
$\phi  $ it would not be possible to compute the required one-step
expectation $PG$ in closed form.

Figure~\ref{f:example6b} shows a typical realization of the results
of the three estimators, for $n=10000$ simulation steps. The
``true'' posterior mean of $V$ is estimated to be $\approx 4.254$
after 10 million simulation steps, and
the corresponding variance reduction factors, estimated from $T=100$
repetitions of the same experiment, are shown in
Table~\ref{tab:ex6}.
\begin{figure}[ht!]
\centerline{\includegraphics[width=4.6in]{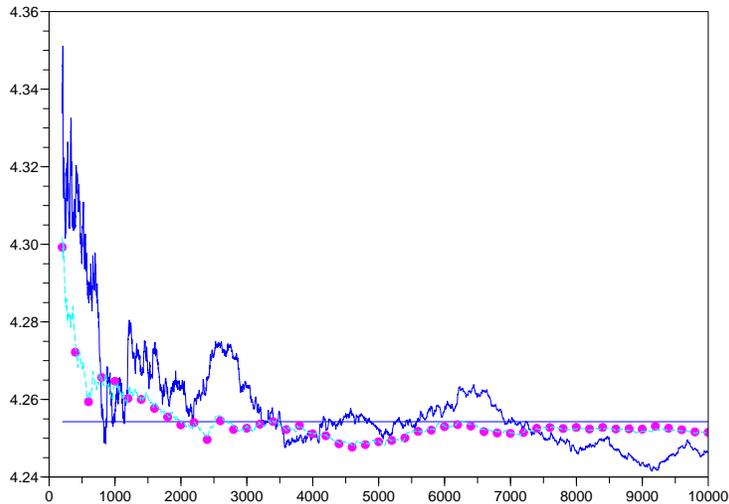}}
\caption{The sequence of the standard ergodic averages
is shown as a solid blue line;
the adaptive estimates
$\mu_{n,\trev,1}(F)$ as bold magenta dots;
and the adaptive estimates
$\mu_{n,\trev,2}(F)$ as a cyan dashed line.
For visual clarity,
the values $\mu_{n,\trev,1}(F)$
are plotted only every 200 simulation
steps.}
\flabel{example6b}
\end{figure}

\begin{table}[ht!]
  \begin{center}
    \begin{tabular}{|c||c|c|c|c|}
      \hline
      \multicolumn{5}{|c|}{\bf Variance reduction factors} \\
      \hline
      & \multicolumn{4}{|c|}{\em Simulation steps}\\
        \hline
{\em Estimator}& $n=10000$& $n=50000$& $n=100000$& $n=200000$\\
     \hline
$\mu_{n,\trev,1}(F)$ & 2.58 & 7.62 & 9.34 & 8.13 \\
     \hline
$\mu_{n,\trev,2}(F)$ & 7.89 & 7.48 & 10.46 & 8.54 \\
     \hline
     \end{tabular}
  \end{center}
     \caption{Estimated factors by which the
    variance of $\mu_n(F)$
    is larger than the corresponding
    variances of
    $\mu_{n,\trev,1}(F)$
    and $\mu_{n,\trev,2}(F)$,
    respectively,
    after
    $n=10000,50000,100000$ and $200000$ simulation steps.}
     \label{tab:ex6}
\end{table}


\section{Variance, Bias and Choosing Between $\mu_{n,\trev,1}$
and $\mu_{n,\trev,2}$}
\label{s:bias}

We examine how the use of the adaptive
estimators estimators $\mu_{n,\trev,1}(F)$ and $\mu_{n,\trev,2}(F)$
can affect the estimation bias, especially in cases where the
initial values of the sampler are far from the true mean of the
target distribution. Also we briefly discuss the different
advantages and disadvantages offered by each of these two
estimators, and conclude that, generally, the preferable choice is
$\mu_{n,\trev,2}(F)$.

\subsection{Estimation bias}

The primary focus of the present
work is on variance reduction, more
specifically, on reducing the
asymptotic variance $\sigma_F^2$
of the estimators $\mu_n(F)$.
This variance
is a ``steady-state'' object,
in that it characterizes the
long-term behavior of the averages
$\mu_n(F)$ and depends
neither on the initial condition
$X_0=x$ nor on the transient behavior
of the chain.
The bias, on the other hand, depends
heavily on the initial condition,
and vanishes asymptotically. Indeed,
according to the expression
in (\ref{eq:sumFhat}) for the
solution $\hat{F}$ of the Poisson
equation (\ref{eq:poisson}),
\be
\mbox{bias}_x(\mu_n(F))
:=
    E_x[\mu_n(F)]-\pi(F)
=
    \frac{1}{n}\sum_{k=0}^{n-1}[P^kF(x)-\pi(F)]
=
    \frac{1}{n}[\hat{F}(x)+o(1)],
    \label{eq:bias1}
\ee
which decays to zero
approximately like $\hat{F}(x)/n$,
as $n\to\infty$.

If instead of the standard ergodic
averages $\mu_n(F)$ we use
an estimator of the form $\mu_n(F_\theta)$
based on a control variate
$U=G-PG$ for some function $G$,
then,
replacing $F$ with $F_\theta$ in the
above computation shows that the bias
of $\mu_n(F_\theta)$ is,
\be
\mbox{bias}_x(\mu_n(F_\theta))
:=
    E_x[\mu_n(F_\theta)]-\pi(F)
= \frac{1}{n}[\hat{F_\theta}(x)+o(1)] =\frac{1}{n}[\hat{F}(x)-\theta
G(x)+o(1)], \label{eq:bias2} \ee where we used the fact that
$\hat{F}_\theta=\hat{F}-\theta G$, as shown in (\ref{eq:FthetaHat}).
Therefore, the function $G$ that minimizes (to first order) the bias
of the estimates $\mu_n(F_\theta)$ is again the solution of the
Poisson equation $G=\hat{F}$. Of course this can also be seen
directly from the definition of $F_\theta$: As in (\ref{eq:zero}),
if $G=\hat{F}$ and $\theta=1$, then, $F-\theta U\equiv \pi(F)$,
leading to an estimator with zero bias and zero variance.

As noted earlier, the solution $\hat{F}$ to
the Poisson equation cannot be computed
in the vast majority of realistic examples
of nontrivial Markov
chains appearing in applications, if for
no other reason, because it requires knowledge
of the mean $\pi(F)$.
Instead, a more pragmatic goal is to try
and choose a ``good'' value for
the parameter $\theta$,
so that the resulting estimator
$\mu_n(F_\theta)$ has
significantly smaller bias than
$\mu_n(F)$.
Unlike the variance,
the bias depends heavily on
the initial condition $x$,
so there is no obvious choice that
makes $\theta G(x)$
``close'' to $\hat{F}(x)$
for all $x$. In fact,
for good variance reduction results
we wish to have $G(x)\approx \hat{F}(x)$
for most values $x$
near the bulk of the target distribution $\pi$,
whereas for the bias we need to have
$G(x)\approx \hat{F}(x)$ at the initial value
$x$ of the chain, which may
well be out in the tail of $\pi$.
Nevertheless,
it may be natural to expect
that taking $\theta\approx\theta^*$
could be a good general substitute.
Although $\theta^*$ does not
eliminate the bias entirely, it does
bring $\theta G$ ``as close as possible''
to $\hat{F}$, where ``closeness'' here
is measured in the sense of minimizing
$\sigma_{F_\theta}^2$.

In order to examine whether the
choice $\theta\approx\theta^*$
does indeed offer an advantage in
terms of the bias,
we revisit Example~2
from~\Section{theta-rev}, where it was observed
that in some cases the adaptive estimators
$\mu_{n,\trev,1}(F)$ and
$\mu_{n,\trev,2}(F)$ did offer
a significant reduction in the bias.

\medskip

\noindent
{\bf Example 2 revisited: Bias and MSE. }
As before, we use the random-scan Gibbs sampler
to simulate from a bivariate Normal vector
$(X,Y)\sim\pi(x,y)$,
where the expected values of both $X$ and $Y$ are
zero, $\VAR(X)=1,$ $\tau^2=\VAR(Y)=10$, and
their covariance $E(XY)=\rho\tau$ with $\rho=0.99.$
To estimate the expected value of $X$ under $\pi$,
we let $F(x,y)=x$, $G(x,y)=x+y$ and
take the control variate $U=G-PG$.

In \Section{theta-rev} it was noted that,
when the initial values of the sampler
were relatively far from their mean
(so that the samples where initially heavily
biased), the adaptive estimators
$\mu_{n,\trev,1}(F)$ and $\mu_{n,\trev,2}(F)$
not only reduced the variance, but also
appeared to be correcting for the estimation
bias; see Figure~\ref{f:example2c}.
This agrees with the intuition obtained
by the discussion following the
computations in (\ref{eq:bias1}) and (\ref{eq:bias2})
above. In order to get a more precise idea of
the effect of the use of the adaptive estimators
$\mu_{n,\trev,1}(F)$ and $\mu_{n,\trev,2}(F)$
on the bias and the overall estimation error,
we estimate the factors by which each of these
two estimators improves
$(i)$~the bias; $(ii)$~the variance;
and $(iii)$~the overall estimation
mean-squared error (MSE).
The results are shown in
Tables~\ref{tab:ex2b} and~\ref{tab:ex2d};
Table~\ref{tab:ex2b} shows simulation
results for a sampler started from
initial values near the true
mean of the distribution, $x_0=y_0=0.1$,
and Table~\ref{tab:ex2d} shows corresponding
results with initial values
$x_0=4$, $y_0=12$.

\begin{table}[ht!]
  \begin{center}
    \begin{tabular}{|c||c|c|c|c|c|}
      \hline
      \multicolumn{6}{|c|}
    {\bf Example 2: $x_0=y_0=0.1$} \\
      \hline
      \hline
      \multicolumn{6}{|c|}
    {\bf Bias reduction factors} \\
        \hline
    {\em Estimator} &$n=1000$ &$n=10000$ &$n=20000$ &$n=50000$ &$n=100000$ \\
     \hline
$\mu_{n,\trev,1}(F)$    & 0.80 & 2.06 & 1.27 & 1.65 & 1.00 \\
     \hline
$\mu_{n,\trev,2}(F)$    & 0.83 & 1.31 & 1.02 & 1.58 & 0.75 \\
     \hline
      \hline
      \multicolumn{6}{|c|}
    {\bf Variance reduction factors} \\
        \hline
$\mu_{n,\trev,1}(F)$    & 2.46 & 7.27 & 8.06 & 8.77 & 9.60 \\
     \hline
$\mu_{n,\trev,2}(F)$    & 3.51 & 6.34 & 6.62 & 8.15 & 9.33 \\
     \hline
      \hline
      \multicolumn{6}{|c|}
    {\bf MSE reduction factors} \\
      \hline
$\mu_{n,\trev,1}(F)$    & 2.45 & 7.26 & 8.04 & 8.64 &  9.54\\
     \hline
$\mu_{n,\trev,2}(F)$    & 3.46 & 6.29 & 6.59 & 8.03 &  9.29\\
     \hline
     \end{tabular}
  \end{center}
     \caption{Estimated factors by which the bias,
	variance, and MSE of $\mu_n(F)$
    is larger than that of
    $\mu_{n,\trev,1}(F)$ and
    $\mu_{n,\trev,2}(F)$,
    respectively, after
    $n=1000,10000,20000,50000$ and 100000 simulation steps.}
     \label{tab:ex2b}
\end{table}

\begin{table}[ht!]
  \begin{center}
    \begin{tabular}{|c||c|c|c|c|c|}
      \hline
      \multicolumn{6}{|c|}
    {\bf Example 2: $x_0=4,\; y_0=12$} \\
      \hline
      \hline
      \multicolumn{6}{|c|}
    {\bf Bias reduction factors} \\
        \hline
    {\em Estimator} &$n=1000$ &$n=10000$ &$n=20000$ &$n=50000$ &$n=100000$ \\
     \hline
$\mu_{n,\trev,1}(F)$    & 1.95 & 3.66 & 4.45 & 7.97  & 7.97\\
     \hline
$\mu_{n,\trev,2}(F)$    & 6.88 & 7.39 & 8.35 & 13.52 & 9.22\\
     \hline
      \hline
      \multicolumn{6}{|c|}
    {\bf Variance reduction factors} \\
        \hline
$\mu_{n,\trev,1}(F)$    & 1.59 & 8.04 & 9.01  & 8.66 & 9.08\\
     \hline
$\mu_{n,\trev,2}(F)$    & 0.89 & 7.00 & 7.91  & 8.72 & 8.72\\
     \hline
      \hline
      \multicolumn{6}{|c|}
    {\bf MSE reduction factors} \\
      \hline
$\mu_{n,\trev,1}(F)$    & 2.57  & 8.34  & 9.75  & 9.11 & 9.34\\
     \hline
$\mu_{n,\trev,2}(F)$    & 2.57  & 7.60  & 9.01  & 9.22 & 8.98\\
     \hline
     \end{tabular}
  \end{center}
     \caption{Estimated factors by which the bias,
	variance, and MSE of $\mu_n(F)$
    is larger than that of
    $\mu_{n,\trev,1}(F)$ and
    $\mu_{n,\trev,2}(F)$,
    respectively, after
    $n=1000,10000,20000,50000$ and 100000 simulation steps.}
     \label{tab:ex2d}
\end{table}

The bias $E_x[\mu_n(F)]-\pi(F)$ of the standard estimators
$\mu_n(F)$ was computed from $T=200$ independent repetitions
of the same experiment, in a way similar to that used for
the variance in the earlier examples; see the discussion
in Example~1. Specifically,
for $\mu_n(F)$, $T=200$ different estimates
$\mu^{(i)}_n(F)$, for $i=1,2,\ldots,T$, were obtained
from $T=200$ independent runs of the Gibbs sampler.
Then the bias of $\mu_n(F)$ was estimated by,
\be
\bar{\mu}_n(F)-\pi(F)\;:=\;
\frac{1}{T}\sum_{i=1}^{T}\mu^{(i)}_n(F)-\pi(F),
\label{eq:red_est_b}
\ee
and the same procedure was applied to estimate the
bias of $\mu_{n,\trev,1}(F)$ and  $\mu_{n,\trev,2}(F)$.
The bias reduction factors shown in the two tables are
the ratios of the corresponding
(absolute values of the) bias estimates.
The variance reduction factors were computed
as before, and the MSE reduction factors
were computed in an analogous manner.

In both cases, the results clearly show that
both estimators
$\mu_{n,\trev,1}(F)$ and  $\mu_{n,\trev,2}(F)$
greatly reduce the estimation error, not only
in terms of their asymptotic variance, but
in terms of the bias and of the overall estimation
error as well.

\subsection{Choosing between the two estimators}
\slabel{choice}

In the simulation examples presented
so far as well as in many more experiments,
we observed that the overall performance
of the two estimators is fairly similar.
One difference is that, in cases where
the initial values of the sampler were very far
from the bulk of the mass of the target
distribution $\pi$, sometimes
$\hat{\theta}_{n,\trev,1}$ converged
faster than $\hat{\theta}_{n,\trev,2}$
and the corresponding estimator
$\mu_{n,\trev,1}(F)$ gave better results
than $\mu_{n,\trev,2}(F)$. The reason for
this discrepancy is the existence of a
the time-lag in the definition of
$\hat{\theta}_{n,\trev,2}$: When
the initial simulation phase produces samples
that approach the area near the mode
of the distribution approximately monotonically,
the denominator of $\hat{\theta}_{n,\trev,2}$
accumulates a systematic one-sided error, and
therefore takes longer to converge. But this
is a transient phenomenon, and can be addressed
(and often eliminated) by including a burn-in phase
in the simulation.

One the other hand, we observed
that the estimator $\hat{\theta}_{n,\trev,2}$
was systematically more stable than
$\hat{\theta}_{n,\trev,1}$, especially in
the more complex MCMC scenarios involving
multiple control variates. This was particularly
pronounced in cases where the denominator
of $\theta^*$ is near zero. There,
because of the inevitable fluctuations
in the estimation of this denominator,
the values of $\hat{\theta}_{n,\trev,1}$
fluctuated wildly between large negative
and positive values, whereas the estimates
$\hat{\theta}_{n,\trev,2}$ were much
more reliable since, by definition, the
denominator of
$\hat{\theta}_{n,\trev,2}$ is always nonnegative.

In conclusion, we find that between
$\mu_{n,\trev,1}$ and
$\mu_{n,\trev,2}$,
the estimator
$\mu_{n,\trev,2}$ is generally the
more reliable, preferable choice.
In all the examples that follow, we will
restrict attention to this estimator;
see also the comments at the end of
\Section{mltp-theory}.


\section{Using Multiple Control Variates Simultaneously}
\label{s:multiple}

\subsection{Adaptive estimators with multiple control variates}
\label{s:mltp-theory}

Starting from the same setting of a Markov chain $\{X_n\}$
with transition kernel $P$, invariant measure $\pi$,
and a function $F:\state\to\RL$ whose mean under $\pi$
is to be estimated, suppose that, instead
of using a single control variate $U=G-PG$, we wish to
use multiple $U_j=G_j-PG_j$, $j=1,2,\ldots,k$. One reason
for such a choice is so that the optimal
$G=\hat{F}$ may potentially be approximated
as a linear combination of ``basis functions'' $G_j$,
namely, $\hat{F}\approx\sum_j\theta_jG_j$.

Formally, let $G:\state\to\RL^k$ denote the column vector
$G=(G_1,G_2,\ldots,G_k)^t$, where each $G_j$ is
a given function $G_j:\state\to\RL$, and similarly
write $U=(U_1,U_2,\ldots,U_k)^t$
for the column vector of control variates
$U_j=G_j-PG_j$.
For any coefficient vector
$\theta=(\theta_1,\theta_2,\ldots,\theta_k)^t\in\RL^k$,
we write $F_\theta=F-\langle\theta,U\rangle$ and consider
the corresponding modified estimator for $\pi(F)$,
$$\mu_n(F_\theta)=\mu_n(F)-\langle\theta,\mu_n(U)\rangle
= \mu_n(F)-\sum_{j=1}^k\theta_j\mu_n(U_j).$$
[Here and throughout the paper
all vectors are column vectors, and
$\langle\cdot,\cdot\rangle$
denotes the usual Euclidean inner product.]
Arguing exactly as in the one-dimensional case,
the asymptotic variance of $F_\theta$ can be expressed as,
\be
\sigma_{F_\theta}^2=\sigma_F^2-2\pi\Big(
\hat{F}\langle\theta,G\rangle-P\hat{F}\langle\theta,PG\rangle
\Big)
+\pi\Big(
\langle\theta,G\rangle^2-
\langle\theta,PG\rangle^2
\Big),
\label{eq:varTheta}
\ee
where, $PG$ stands for
the vector $(PG_1,PG_2,\ldots,PG_k)^t$.

To find the optimal $\theta^*$, differentiate
the quadratic
$\sigma_{F_\theta}^2$ with respect to each $\theta_i$
and set the derivative equal to zero, to obtain,
in matrix notation,
$$\Gamma(G)\theta^*=\pi(\hat{F}G-(P\hat{F})(PG)),$$
where the $k\times k$ matrix $\Gamma(G)$ has entries,
$\Gamma(G)_{ij}=\pi(G_iG_j-(PG_i)(PG_j))$. Therefore,
\be
\theta^*=
\Gamma(G)^{-1}\pi(\hat{F}G-(P\hat{F})(PG)),
\label{eq:thetaStar-k}
\ee
as long as $\Gamma(G)$ is invertible.
Note that this expression is perfectly analogous
to the one-dimensional formula for $\theta^*$
in (\ref{eq:thetaStar}).
Also, in view of equation (\ref{eq:cov})
from Section~\ref{s:cv2},
the entries of $\Gamma(G)$
can be expressed as,
$$\Gamma(G)_{ij}=\pi(G_iG_j-(PG_i)(PG_j))
=\pi(\hat{U_i}G_j-(P\hat{U}_i)(PG_j))
=\sum_{n=-\infty}^\infty\COV_\pi(U_i(X_0),G_j(X_n)).
$$
This shows that $\Gamma(G)$ has the structure
of a covariance matrix and, in particular,
it suggests that $\Gamma(G)$ should be positive
semidefinite. Indeed, the following lemma states
that the entries of $\Gamma(G)$ can be written
in a way which makes both of these assertions
obvious:

\medskip

\noindent
{\bf Lemma 2. } Let $\K(G)$ denote the covariance
matrix of the random variables
$$Y_i:=G_i(X_1)-PG_i(X_0),
\;\;\;\;
i=1,2,\ldots,k,$$
where $X_0\sim \pi$. Then $\Gamma(G)=\K(G)$,
that is,
for all $1\leq i,j\leq k$,
\be
\pi(G_iG_j-(PG_i)(PG_j)) =
    E_\pi\Big[
    \Big(G_i(X_1)-PG_i(X_0)\Big)
    \Big(G_j(X_1)-PG_j(X_0)\Big)
    \Big].
\label{eq:lemma2}
\ee

\noindent
{\sc Proof. }
Expanding the right-hand side of (\ref{eq:lemma2})
we obtain,
$$  \pi(G_iG_j)
    -E_\pi[G_i(X_1)PG_j(X_0)]
    -E_\pi[G_j(X_1)PG_i(X_0)]
    +\pi((PG_i)(PG_j)),
$$
and the result follows upon noting that the
second and third terms above are both equal to
the fourth. To see this, observe that the
second term can be rewritten as,
\ben
    E_\pi\Big\{E\Big[G_i(X_1)PG_j(X_0)\,\Big|\,X_0\Big]\Big\}
=
    E_\pi\Big[E[G_i(X_1)\,|\,X_0]PG_j(X_0)\Big]
=
    \pi((PG_i)(PG_j)),
\een
and similarly for the third term.
\qed

Therefore, the optimal coefficient
$\theta^*$ can also be expressed as,
\be
\theta^*=
\K(G)^{-1}\pi(\hat{F}G-(P\hat{F})(PG)).
\label{eq:thetaStar-k_2}
\ee

Proceeding exactly as before,
for a reversible chain,
starting from the expressions
for $\theta^*$ in
(\ref{eq:thetaStar-k})
and (\ref{eq:thetaStar-k_2})
we obtain:

\medskip

\noindent
{\bf Proposition 2. } If the chain $\{X_n\}$ is
reversible, then the optimal coefficient vector
$\theta^*$ for the control variates
$U_i=G_i-PG_i$, $i=1,2,\ldots,k$ can be expressed
as,
\be
\theta^*=
\theta^*_\trev:=\Gamma(G)^{-1}
\pi\big((F-\pi(F))(G+PG)\big),
\label{eq:theta_rev_k}
\ee
or, alternatively,
\be
\theta^*=
\theta^*_\trev:=\K(G)^{-1}
\pi\big((F-\pi(F))(G+PG)\big).
\label{eq:theta_rev_k_2}
\ee

The proof of Proposition~2 is perfectly analogous
to that of Proposition~1 in the case of a single
control variate.

As before, the expressions
(\ref{eq:theta_rev_k})
and
(\ref{eq:theta_rev_k_2})
suggest estimating
$\theta^*$ via,
\ben
\hat{\theta}_{n,\Gamma}
&=&
    \Gamma_n(G)^{-1}
    [\mu_n(F(G+PG))-
    \mu_n(F)\mu_n(G+PG)]\\
\mbox{or}
\;\;\;\;
\hat{\theta}_{n,\K}
&=&
    \K_n(G)^{-1}
    [\mu_n(F(G+PG))-
    \mu_n(F)\mu_n(G+PG)],
\een
where the $k\times k$
matrices $\Gamma_n(G)$ and $\K_n(G)$ are
defined, respectively, by,
\ben
(\Gamma_n(G))_{ij}
&=&
    \mu_n(G_iG_j)-\mu_n((PG_i)(PG_j))\\
\mbox{and}
\;\;\;\;
(\K_n(G))_{ij}
&=&
\frac{1}{n}\sum_{t=0}^{n-1}
(G_i(X_t)-PG_i(X_{t-1}))
(G_j(X_t)-PG_j(X_{t-1})).
\een
The resulting estimators,
$\mu_n(F_{\hat{\theta}_{n,\Gamma}})$
and $\mu_n(F_{\hat{\theta}_{n,\K}})$
for $\pi(F)$
based on the vector of
control variates $U=G-PG$
and the coefficients
$\hat{\theta}_{n,\Gamma}$
and
$\hat{\theta}_{n,\K}$,
respectively,
are defined analogously to the
single-control-variate case as,
\be
\mu_{n,\Gamma}(F)
&:=&
    \mu_n(F_{\hat{\theta}_{n,\Gamma}})
    \;=\;\mu_n(F)-\langle\hat{\theta}_{n,\K},\mu_n(U)\rangle
    \label{eq:multi-est1}\\
\mbox{and}
\;\;\;\;
\mu_{n,\K}(F)
&:=&
    \mu_n(F_{\hat{\theta}_{n,\Gamma}})
    \;=\;\mu_n(F)-\langle\hat{\theta}_{n,\K},\mu_n(U)\rangle.
    \label{eq:multi-est2}
\ee

Recall that in~\Section{choice} we concluded
that, for the case of a single control variate,
the adaptive estimator $\mu_{n,\trev,2}$ was
generally preferable to $\mu_{n,\trev,1}$.
For the same reasons, and also based on the
results of extensive simulation experiments
with multiple control variates,
we similarly conclude that $\mu_{n,\K}(F)$
is more reliable, more stable,
and generally preferable to $\mu_{n,\Gamma}(F)$.
Therefore, in all of our subsequent examples
we restrict attention to the estimator
$\mu_{n,\K}(F)$.

\subsection{Examples}

Here we re-examine two of the earlier
examples, and illustrate how the use
of multiple control variates can
often provide a much greater
improvement in estimation accuracy.

\medskip

\noindent
{\bf Example 2 revisited. }
Let $(X,Y)\sim\pi(x,y)$ be
a zero mean, bivariate Normal distribution,
with $\VAR(X)=1$, $\VAR(Y)=\tau^2$, and
$E(XY)=\rho\tau$ for some $\rho\in(-1,1)$.
To estimate the expected value of
$X$ under $\pi$ we sample from $\pi$ using
a random-scan Gibbs sampler and set $F(x,y)=x.$
Instead of the single control variate $U=G-PG$
based on $G(x,y)=x+y$, here we consider
two control variates $U_1,U_2$ defined
in terms of $G_1(x,y)=x$ and $G_2(x,y)=y$,
respectively. We examine the performance
of the adaptive estimator
$\mu_{n,\K}(F)$,
and compare it with
the performance of obtained earlier
by the single-control-variate estimator
$\mu_{n,\trev,2}(F)$.

Figure~\ref{f:example2_multi}
depicts a typical realization
of the sequence of estimates
of the standard ergodic averages
$\mu_n(F)$, as well as the corresponding
estimates obtained by
$\mu_{n,\K}(F)$, for $n=20000$
simulation steps.
The parameter values are
$\rho=0.99$ and $\tau^2=10$,
with initial values $x_0=y_0=0.1$.
Table~\ref{tab:ex2_multi}
shows the corresponding variance reduction factors,
estimated from $T=200$ repetitions of the same experiment.

\begin{figure}[ht!]
\centerline{\includegraphics[width=4.4in]{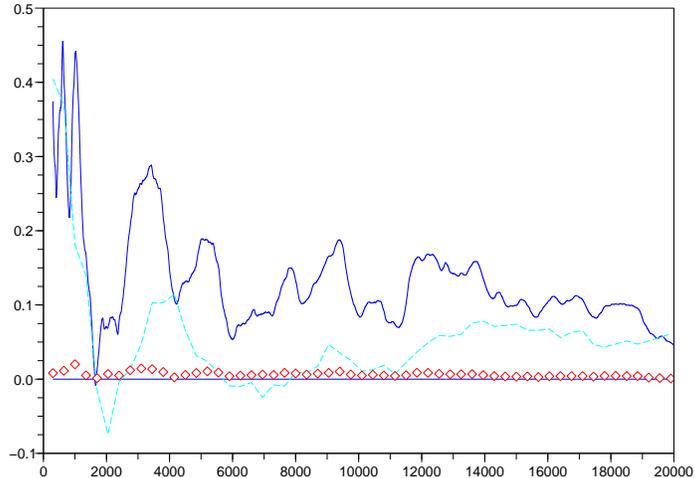}}
\caption{The sequence of the standard ergodic averages
is shown as a solid blue line;
the adaptive estimates
$\mu_{n,\trev,2}(F)$ with respect
to the single control variate $U$
as a cyan dashed line,
and the adaptive estimates
$\mu_{n,\K}(F)$
with respect to the two control
variates $U_1,U_2$ as red diamonds.
For visual clarity,
the values $\mu_{n,\K}(F)$
are plotted only every 350 simulation
steps.}
\flabel{example2_multi}
\end{figure}

\begin{table}[ht!]
  \begin{center}
    \begin{tabular}{|c||c|c|c|c|c|}
      \hline
      \multicolumn{6}{|c|}{\bf Variance reduction factors} \\
      \hline
      & \multicolumn{5}{|c|}{\em Simulation steps}\\
        \hline
{\em Estimator}& $n=1000$& $n=10000$& $n=50000$& $n=100000$& $n=200000$\\
     \hline
$\mu_{n,\trev,2}(F)$    & 2.89 & 6.17  & 8.17 & 7.42 & 9.96 \\
     \hline
$\mu_{n,\K}(F)$     & 4.13 & 27.91 & 122.4 & 262.5 & 445.0 \\
     \hline
     \end{tabular}
  \end{center}
     \caption{Estimated factors by which the
    variance of $\mu_n(F)$
    is larger than the corresponding
    variances of
    $\mu_{n,\trev,2}(F)$ and
    $\mu_{n,\K}(F)$,
    respectively,
    after
    $n=1000,10000,50000,100000$ and $200000$ simulation
    steps.}
     \label{tab:ex2_multi}
\end{table}

Clearly the estimator based
on the two control variates
is extremely effective, and
certainly significantly better
than the one based on a single
control variate. As in Example~4,
this effectiveness is actually
explained by the fact that
the exact solution of
the Poisson equation
in this case is of the
form $\hat{F}(x,y)=ax+by$. Indeed,
it is a simple matter to verify
that,
$\hat{F}(x,y) = \frac{2}{1-\rho^2}[x+\frac{\rho}{\tau}y].$

\medskip

\noindent
{\bf Example 6 revisited. }
Recall the setting of the inference problem in
Example~6 above, where, based on $N=500$
independent observations
$x=(x_1,x_2,\ldots,x_N)$
generated from the mixture
$pN(\mu_1,\sigma^2)+(1-p)N(\mu_2,\sigma^2)$,
we wish to estimate $\mu_1$.
The mixing proportion $p=0.7$ and the variance
$\sigma^2=1$ are fixed and known,
$N(0,10\sigma^2)$ priors
are placed on $\mu_1,\mu_2$,
and each of the binary latent variables
$(Z_i)$ equals $1$ if the corresponding $x_i$
is generated from the first component
of the mixture, and $Z_i=0$ otherwise.
We use the random-scan Gibbs sampler,
based on the full conditionals of the
posterior, given in
(\ref{eq:mu_full})
and (\ref{eq:qi}).

In Section~\ref{sec:simple}, letting $F(\mu_1,\mu_2,Z)=\mu_1$
and using a control variate $U=G-PG$ in terms of the
function $G(\mu_1,\mu_2,Z)=(Np+1/10)\mu_1+\sum_iZ_ix_i,$
we obtained variance reduction factors around 20.
The natural next step is to repeat the same
experiment, this time with two control variates
$U_1,U_2$ defined in terms of
the functions $G_1(\mu_1,\mu_2,Z)=\mu_1$
and $G_2(\mu_1,\mu_2,Z)=\sum_iZ_ix_i$.
In numerous simulation experiments we observed
that, using two control variates in
this case offered no apparent
performance improvement.
This suggests that the ratio of the
coefficients of the functions
$G_2$
and $G_1$,
which was earlier chosen as $1/(Np+0.1)$
based on a heuristic computation,
must be near-optimal. Indeed, after
two million Gibbs steps,
the estimated value of the optimal
parameter vector $\theta^*$ for
the two control variates $U_1,U_2$
was $\approx(3.832,0.0129)$.
The resulting optimal ratio
$0.0123/3.832 \approx 0.0034$
is, as expected,
quite close to
$1/(Np+0.1)\approx 0.0029$.

Next we consider 
using four control variates,
defined in terms of the functions
$G_1,G_2$ above together with
$G_3(\mu_1,\mu_2,Z)=\mu_2$ and
$G_4(\mu_1,\mu_2,Z)=\mu_1^2$.
In this case,
the corresponding variance reduction factors,
estimated from $T=100$ repetitions of the same experiment
(with initial values for the sampler $\mu_1=0$, $\mu_2=1$),
are {\bf 97.47, 138.39, 91.84} and  {\bf 103},
after $n=10000,50000,100000$ and $200000$ simulation
steps, respectively.
Compared to the earlier results
(variance reduction factors around 20),
these results
clearly demonstrate the significant
improvement in estimation accuracy
due to the
simultaneous
use of multiple control
variates.


\section{Four More Complex MCMC Examples}
\slabel{complex}

This section illustrates our proposed methodology applied
to a series of real Bayesian inference problems, providing 
guidelines on how functions $G$ can be chosen
for the construction of effective control variates. 
The first example is a binary probit
model, an early success of MCMC inference through data augmentation;
see \citet{albert-chib:93}. The second example is a simple finite
mixture of normals, another early application of 
data augmentation via Gibbs sampling; see \citet{diebolt-robert:94}. 
What makes this problem particularly interesting is the fact that,
although we impose an {\em a priori} restriction on the
ordering of means, the control variates methodology can
still be applied after a first phase of
{\em unrestricted} MCMC sampling, 
and after the sample has been ordered at
the post-processing stage. If the objective
is to estimate the means, the calculation of 
effective control variates $U$ is still
possible, despite the fact that the 
resulting Markov chain has a particularly 
complex structure. The third example is of 
a Bayesian model-determination problem, in which model searching is
achieved by a discrete Metropolis algorithm on the space 
of candidate models. Such applications have recently 
found tremendous interest, especially in the context
of genetics (see, e.g., \citet{botric}) where the model 
space is endowed with a multimodal discrete
density.  Finally, in the case of a simple log-linear
model we show that, even when we are forced to
consider functions $G$ that are very different from $F$, 
the resulting control variates $U$ can still be very
effective in terms of variance reduction.

\subsection{A binary probit example}

Probit models are a well-known and commonly used class of discrete
regression models; see, for example, the monograph by
\citet{johnson-albert:book} and the references therein. Here we
illustrate the use of the control variate methodology 
when a random-scan Gibbs sampler is used for Bayesian inference
from the posterior of a binary probit model.

Specifically, we begin with
an $N\times k$ matrix
$x=(x_1^t,x_2^t,\ldots,x_N^t)^t$
of known covariates, where each
$x_i$ is a column vector in
$\RL^k$. We also have and a vector
$Y=(Y_1,Y_2,\ldots,Y_N)$ of known
binary responses $Y_i$,
where we assume that the $Y_i$
have,
$$p_i:=\PR\{Y_i=1\}=1-\PR\{Y_i=0\}
= \Phi(x_i^t\beta),
\;\;\;\;i=1,2,\ldots,N,$$
and the unknown parameter vector
$\beta\in\RL^k$ is to be estimated.
To facilitate sampling from the
posterior of $\beta$,
\citet{albert-chib:93}
introduce independent latent random variables
$Z=(Z_1,Z_2,\ldots,Z_N)^t$,
where each $Z_i\sim N(x_i^t\beta,1)$.
In other words $Z=x\beta+\epsilon$, where
$\epsilon\sim N(0,I)$ is independent
noise. Then the $Y_i$ can be expressed,
$Y_i=\IND_{\{Z_i>0\}}$, so that, again,
$p_i=\Phi(x_i^t\beta)$.

If we place a diffuse prior on $\beta$, then
$\pi(\beta|x,Y,Z)\sim N((x^tx)^{-1}(x^tZ),\,(x^tx)^{-1}),$
and the $Z_i$ are conditionally independent given $x,Y,\beta$, with,
\ben
\pi(Z_i|x,Y,\beta)
&\sim&
     N(x_i^t\beta,1)\;\;\;\mbox{conditional on $Z_i>0$, if $Y_i=1$};\\
\pi(Z_i|x,Y,\beta)
&\sim&
     N(x_i^t\beta,1)\;\;\;\mbox{conditional on $Z_i<0$, if $Y_i=0$}.
\een

We consider a specific example using the ``statistics class'' data
from \citet[p.~77]{johnson-albert:book}.
In this case, for $N=30$ students, each $Y_i$ is the indicator of
weather student $i$ passed or failed in a statistics class. There
are $k=2$ covariates $(x_{i1},x_{i2})$ for each student, where
$x_{i1}=1$ for all $i$ and $x_{i2}$ is the $i$th student's SAT Math
test score. We place a diffuse prior on the coefficient vector
$\beta=(\beta_0,\beta_1)$, and we consider the problem of estimating
the posterior mean of $\beta_1$. (The
    parameter $\beta_1$ is chosen as the more
    interesting of the two; the results are very
    similar for the case of $\beta_0$.)
To that end, we let $F(\beta,Z)=\beta_1$,
and we also consider a vector $U=G-PG$ of
control variates based on
five-component function $G$,
$$G(\beta,Z)=(\beta,(x^tx)^{-1}x^tZ,\beta_1^2)^t.$$
For the initial condition of $\beta$ in the
sampler we took its least-squares estimate
$\hat{\beta}:=(x^tx)^{-1}(x^ty)$,
and for $Z$ we simply drew a sample
from its full conditional density as above.
The choice of the function $G$ for the construction
of control variates is pretty self-evident:
The variables $\beta_0,\beta_1$ and $\beta_1^2$
should obviously be strongly correlated with the
target variable $\beta_1$, and
the vector $(x^tx)^{-1}x^tZ$ is included
in an attempt to minimize the effect
of the mean of $\beta$ under its full conditional.

The result of a typical realization of the random-scan Gibbs sampler
after $15000$ iterations is shown in Figure~\ref{f:binary_probit}.
The horizontal line shows the ``true'' value of $\pi(F)
\approx 0.03759$, the result of $\mu_n(F)$ after 10
million Gibbs iterations. The variance reduction
factors obtained by $\mu_{n,\K}(F)$,
estimated after $T=100$ repetitions of this experiment,
are {\bf 5.08, 34.22, 53.54, 88.37} and {\bf 69.72}, after
$n=1000$, $10000$, $50000$, $100000$ and $200000$
iterations, respectively.

\begin{figure}[ht!]
\centerline{\includegraphics[width=4.6in]{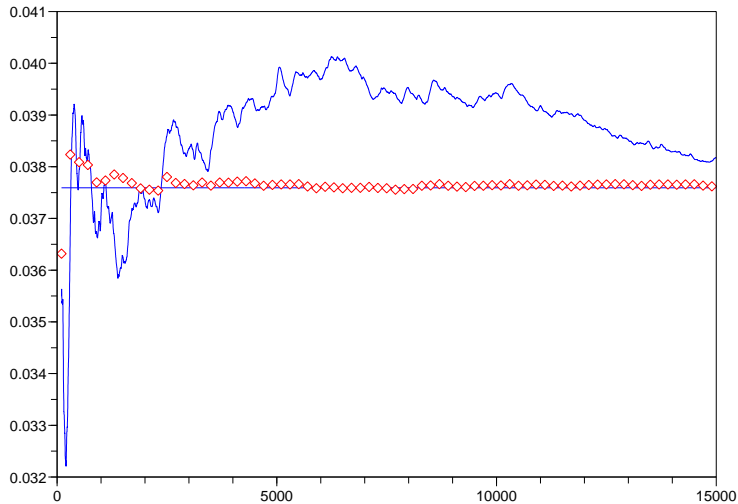}}
\caption{Probit: The sequence of the standard ergodic averages
is shown as a solid blue line and
the adaptive estimates
$\mu_{n,\K}(F)$
as red diamonds.
The adaptive estimates
are plotted only every 200 simulation
steps.}
\flabel{binary_probit}
\end{figure}


\subsection{Gaussian mixtures}
\label{s:Gcomplex}
Mixtures of densities provide a versatile class of statistical
models that have received a lot of attention from both a theoretical
and a practical perspective, for many decades now.
Mixtures primarily serve as a means of modelling heterogeneity for
classification and discrimination, and as a way of formulating
flexible models for density estimation. Although one of the fist
major success stories in the MCMC community was the Bayesian
implementation of the finite Gaussian mixtures problem,
\citet{tanwon87,diebolt-robert:94}, there are still numerous unresolved
issues in inference for finite mixtures, as discussed, for example,
in the recent review paper by \citet{Marin05}. These difficulties
emanate primarily from the fact that such models are often ill-posed
or non-identifiable. In terms of Bayesian inference via MCMC, these
issues reflect important problems in prior specifications and label
switching. In particular, improper priors are hard to use and proper
mixing over all posterior modes requires enforcing label-switching
moves through Metropolis steps. Detailed discussions of the dangers
emerging from prior specifications and identifiability constraints
can be found in \citet{Marin05,Lee08,Jasra05}.

Here we generalize the estimation setting of Example~6 above,
by employing a more realistic two-component
Gaussian mixtures model as follows.
Starting with $N=500$ data points $x=(x_1,x_2,\ldots,x_N)$
generated from the mixture distribution
$0.7N(0,0.5^2)+0.3N(0.1,3^2)$, and assuming that the
means, variances and mixing proportions are all unknown,
we consider the problem of estimating the
two means. The usual Bayesian formulation enriches
that of Example~6 by introducing parameters
$(\mu_1, \mu_2, \sigma_1,\sigma_2,Z,p)$, as follows.
The data are assumed to be i.i.d.\ from
$pN(\mu_1,\sigma_1^2)+(1-p)N(\mu_2,\sigma_2^2)$,
and we place the following priors:
$p \sim \mbox{Dirichlet}(\delta,\delta)$,
the two means $\mu_1,\mu_2$ are independent
with each $\mu_j \sim N(\xi,\kappa^{-1})$,
and similarly the variances are independent
with each $\sigma_j^{-2} \sim \mbox{Gamma}(\alpha,\beta)$.
We adopt the vague, data-dependent
prior structure of \citet{rich}: We set
$\delta=1$, we let $\xi$ equal to the empirical mean
of the data $x$, $\kappa^{-1/2}$ is taken
to be equal to the data range, $\alpha=2$ and $\beta=0.02\kappa^{-1}$.
As before, conditional on the parameters
$(\mu_1, \mu_2, \sigma_1,\sigma_2,p)$,
the latent variables $Z=(Z_1,Z_2,\ldots,Z_N)$
are i.i.d.\ with $\Pr\{Z_i=1\}=1-\Pr\{Z_i=2\}=p,$
and, given the entire parameter vector
$(\mu_1, \mu_2, \sigma_1,\sigma_2,Z,p)$,
the data $x=(x_i)$ are i.i.d.\
with each $x_i$ having distribution
$N(\mu_j,\sigma_j^2)$ if $Z_i=j$,
for $i=1,2,\ldots,N$, $j=1,2$.

In order to estimate the mean vector $(\mu_1,\mu_2)$ we sample from
the posterior via a standard random-scan Gibbs sampler, and we also
introduce the {\em a priori} restriction that $\mu_1 < \mu_2$. In terms of
the sampling itself, as noted by \citet{Stephens97}, it is
preferable to first obtain draws from the unconstrained posterior
distribution and then to impose the identifiability constraint at
the post-processing stage. In each iteration, the random-scan Gibbs
sampler selects one of the four parameter blocks $(\mu_1,\mu_2)$,
$(\sigma_1,\sigma_2)$, $Z$ or $p$, each with probability $1/4$, and
draws a sample from the corresponding full conditional density.
These densities are all of standard form and easy to sample from;
see, for example, \citet{rich}. In particular, the two means are
conditionally independent with each, \be \mu_j\sim N\left(
\frac{\sigma_j^{-2}\sum_i(y_i+\kappa\xi)\IND_{\{Z_i=j\}}}
{\sigma_j^{-2}n_j+\kappa}, \frac{1}{\sigma_j^{-2}n_j+\kappa}
\right), \label{eq:fullCD} \ee where $n_j=\#\{i:Z_i=j\}$, for
$j=1,2$. Note that the data $x$ have been generated so that the two
means are very close, which results in frequent label switching
throughout the MCMC run and in near-identical marginal densities of
$\mu_1$ and $\mu_2$.

We perform a post-processing relabelling of the sampled values
according to the above restriction,
and we denote the ordered sampled vector by
$(\mu^o_1, \mu^o_2, \sigma^o_1,\sigma^o_2,Z^o,p^o)$.
In order to estimate the posterior mean
of the smaller of the two means, we let,
$$F(\mu_1, \mu_2, \sigma_1,\sigma_2,Z,p)
:=\mu_1^o=\min\{\mu_1,\mu_2\}.$$
To reduce the variance of the
estimator $\mu_n(F)$ we consider a bivariate control variate
$U=G-PG$, where the function $G=(G_1,G_2)^t$ is
selected as follows. For $G_1$ we take the obvious
choice, $G_1(\mu_1, \mu_2, \sigma_1,\sigma_2,Z,p)=
\mu_1^o$, so that, $PG_1$, the expected value
of $\min\{\mu_1,\mu_2\}$ under (\ref{eq:fullCD}),
is easily seen to be,
\be
&&
PG_1(\mu_1, \mu_2, \sigma_1,\sigma_2,Z,p)
\;=\;
    \frac{3}{4}G_1(\mu_1, \mu_2, \sigma_1,\sigma_2,Z,p)
    \nonumber\\
&& \hspace{0.8in}+\;
\frac{\nu_1}{4}\Phi\Big(\frac{\nu_2-\nu_1}{\sqrt{\tau_1^2+\tau_2^2}}\Big)
+\frac{\nu_2}{4}\Phi\Big(\frac{\nu_1-\nu_2}{\sqrt{\tau_1^2+\tau_2^2}}\Big)
-\frac{1}{4}\sqrt{\tau_1^2+\tau_2^2}
\phi\Big(\frac{\nu_2-\nu_1}{\sqrt{\tau_1^2+\tau_2^2}}\Big),
\hspace{0.3in} \label{eq:PG1} \ee where $\nu_j$ and $\tau_j^2$ are
the means and variances of $\mu_j$, respectively, for $j=1,2$, under
the full conditional densities in (\ref{eq:fullCD}); see, for
example, \citet{cain94}. Clearly this introduces a significant
amount of unwanted variability in $U_1=G_1-PG_1$, so, in order to
cancel it out, we choose $G_2$ to approximately cancel out the last
three terms of the above expression. Since the nonlinear terms
involving $\phi$ and $\Phi$ are hard to handle analytically and are
also bounded, and since we expect the dependence on the mean vector
to be taken care of by $G_1$, we focus on approximating the
$\sqrt{\tau_1^2+\tau_2^2}$ factor. Since $\kappa$ will be typically
small  compared to $n_1$ and $n_2$, we approximate $\tau_1^2$ by
$\sigma_1^2/(Np)$ and $\tau_2^2$ by $\sigma_2^2/(N(1-p))$. And since
we expect the influence of $\sigma_1^o$ to be dominant over that of
$\sigma_2^o$ with respect to $\mu_1^o$, a straightforward
first-order Taylor expansion shows that the dominant linear term is
$\sigma_1^o$, suggesting the choice $G_2(\mu_1, \mu_2,
\sigma_1,\sigma_2,Z,p)=\sigma_1^o$.

To compute $PG_2$, we first calculate
the probability $p(\mbox{order})$ that
$\mu_1<\mu_2$ under (\ref{eq:fullCD}),
$$p(\mbox{order})
= \frac{\Phi \big(E( \mu_2|\cdots )- E(\mu_1|\cdots) \big)}
{\sqrt{E(\sigma^2_1|\cdots ) + E(\sigma^2_2|\cdots)}},$$ where all
four expectations above are taken under the corresponding full
conditional densities, and, since the full conditional of each
$\sigma^{-2}_j$ is a Gamma density, the expectations of $\sigma_1$,
$\sigma_2$, $\sigma_1^2$, and $\sigma_2^2$, are all available in
closed form. Therefore, $p(\mbox{order})$ can be computed
explicitly, and, \ben &&
    PG_2(\mu_1, \mu_2, \sigma_1,\sigma_2,Z,p)
    \;=\;\frac{1}{2}G_2(\mu_1, \mu_2, \sigma_1,\sigma_2,Z,p)\\
&&
    \hspace{0.6in}
    +\; \frac{1}{4} \left[
    \IND_{\{\mu_1<\mu_2\}}
    E(\sigma_1|\cdots)
    +\IND_{\{\mu_1>\mu_2\}}E(\sigma_2|\cdots) \right]
    +
    \frac{1}{4} \Big[ p(\mbox{order}) \sigma_1 +
    (1-p(\mbox{order})) \sigma_2 \Big],
\een
where, again, the expectations are taken under the
corresponding full conditional densities.

With this choice for $G=(G_1,G_2)^t$,
the variance reduction factors obtained
by $\mu_{n,\K}(F)$
(estimated from $T=100$ repetitions)
are {\bf 16.17, 25.36, 38.99, 44.5} and  {\bf 36.16},
after $n=1000, 10000,50000,100000$ and $200000$ simulation steps,
respectively.
Figure~\ref{f:mixtures} shows 
the results of a
typical simulation run.
The initial values of the sampler were taken
after a 1000-iteration burn-in period, and
the horizontal line in the graph depicting
the ``true'' value of the posterior mean
of $F$ was obtained after 5 million Gibbs iterations.

\begin{figure}[ht!]
\centerline{\includegraphics[width=4.6in]{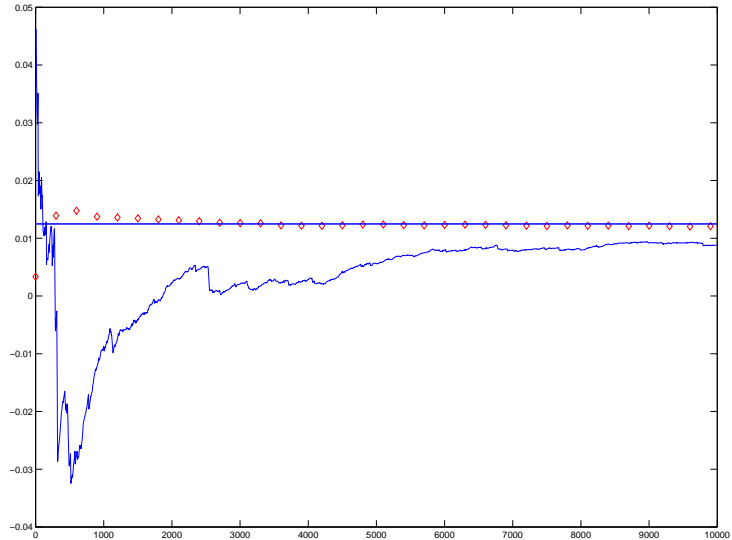}}
\caption{Two-component Gaussian mixture model: The sequence of the standard
ergodic averages for $\mu_1$ is shown as a solid blue line and the
adaptive estimates $\mu_{n,\K}(F)$,
reported every $300$ iterations,
as red diamonds. } \flabel{mixtures}
\end{figure}


\subsection{A two-threshold autoregressive model}
\label{s:setar}
We revisit the monthly U.S.\ 3-month treasury bill rates,
from January $1962$ until December $1999$,
previously analyzed by \citet{ddh} using flexible
volatility threshold models.  The time series has $N=456$ points
and is denoted by $r=(r_t)=(r_t\;;\;t=1,2,\ldots,N)$.
Here we model these
data in terms of a self-exciting threshold autoregressive
model, with two regimes; it is one of the models proposed
by
\citet{Pfann}, and it is defined as,
\begin{equation}
\Delta r_t = \left\{
\begin{array}{ll}
\alpha_{10} + \alpha_{11} r_{t-1} & r_{t-1} < c_1 \\
\alpha_{20} + \alpha_{21} r_{t-1} & r_{t-1} \geq c_1
\end{array} \right\} + \left\{
\begin{array}{ll}
\sigma_1 \epsilon_t & r_{t-1} < c_2 \\
\sigma_2 \epsilon_t & r_{t-1} \geq c_2
\end{array} \right\},
\label{setar}
\end{equation}
where $\Delta r_t = r_t - r_{t-1}$, and
the parameters $c_1$, $c_2$ are the thresholds
where mean or volatility regime shifts occur.
Instead, we re-write the model as,
\begin{equation}
\Delta r_t =
\left\{
\begin{array}{ll}
\alpha_{10} + \alpha_{11} r_{t-1} & r_{t-1} < c_1 \\
\alpha_{20} + \alpha_{21} r_{t-1} & r_{t-1} \geq c_1
\end{array} \right\} + \left\{
\begin{array}{ll}
\sigma \epsilon_t & r_{t-1} < c_2 \\
\sigma(1+\gamma)^{1/2} \epsilon_t & r_{t-1} \geq c_2
\end{array} \right\},
\label{setar2}
\end{equation}
where $\gamma \geq -1$ characterizes the jump in $\sigma^2$ between
the two volatility regimes. Whereas \citet{Pfann} use a Gibbs
sampler to estimate the parameters of the model in (\ref{setar}), we
exploit the parameterization (\ref{setar2}) as follows.  We adopt
independent improper conjugate priors for the variance,
$\pi(\sigma^2) \propto \sigma^{-2}$, and for the regression
coefficients, $\pi(\alpha_{ij}) \propto 1$. We take the prior for
each of $c_1$ and $c_2$ to be a discrete uniform over the distinct
values of $\{r_t\}$, except the two smallest and largest values of
$\{r_t\}$ so that identifiability is obtained; and the prior for
$\gamma$ to be an exponential density with mean one,
shifted to $-1$.

Our goal is to estimate the posterior
probability $\pi(c_1^1,c_2^1|r)$
of the most likely model, that is, 
of the model corresponding to the pair 
of thresholds $(c_1^1,c_2^1)$ maximizing
$\pi(c_1,c_2|r)$.
In the above formulation, (\ref{setar2}) 
can be written equivalently as,
$
R = X \alpha + \epsilon,
$
where $R = (\Delta r_2, \ldots, \Delta r_T)^t$, $\alpha =
(\alpha_{10}, \alpha_{11}, \alpha_{20}, \alpha_{21})$, $\epsilon$
is a zero-mean Gaussian vector with covariance matrix
$\Sigma,$ and $X$ is the design matrix with row $t$
given by (1 $r_{t-1}$ 0 0), if $r_{t-1} \leq c_1$, and by (0 0 1
$r_{t-1}$) otherwise. The covariance matrix of the errors, $\Sigma$,
is diagonal with $\Sigma_{tt}=\sigma^2$ if
$r_{t-1} \leq c_2$, and $\Sigma_{tt}=(1+\gamma)\sigma^2$, otherwise.
Integrating out the parameters $\alpha$ and $\sigma$,
the marginal likelihood
of the data $r$ with known $c_1$, $c_2$ and $\gamma$
is,
$$
p(r | \gamma,c_1,c_2) \propto \exp\Big\{ -\frac{1}{2} \Big[ |\Sigma|
+ \log |X^t \Sigma^{-1} X| + N \log (R^t \Sigma^{-1} R -
\hat{\alpha}^t X^t \Sigma^{-1} X \hat{\alpha}) \Big]\Big\}, $$ where
$\hat{\alpha}=(X^tX)^{-1} X^t R$ is the least-squares estimate of
$\alpha$; see, for example, \citet{OHFOR}. After further performing
a one-dimensional numerical integration over $\gamma$ by numerical
quadrature, we can write the marginal posterior distribution of
$(c_1,c_2)$ explicitly as $\pi(c_1,c_2|r) \propto p(r|c_1,c_2)$.
Therefore, we can sample from the posterior of the thresholds
$(c_1,c_2)$ by employing a discrete Metropolis-Hastings algorithm on
$(c_1,c_2)$, where the thresholds $c_1,c_2$ take values on the
lattice of all the observed values of the rates $(r_t)$ except the
two farthermost at each end. This way, we replace the
$8$-dimensional Gibbs sampler of \citet{Pfann} for
(\ref{setar}), by a five-dimensional analytical integration over
$\alpha$ and $\sigma$, a numerical integration over $\gamma$, 
and a Metropolis-Hastings algorithm over $(c_1,c_2)$.

Note that this algorithm is computationally less
expensive, and also more reliable since Gibbs
sampling across a discrete and continuous product space 
may encounter `sticky patches' in the parameter space.
The discrete Metropolis-Hastings sampler we employ is based on
symmetric random walk steps,
with vertical or horizontal increments
of size up to ten, over the lattice of all possible values. 
In other words,
the proposed pair $(c'_1,c'_2)$ given the current values $(c_1,c_2)$
is one of the forty ``neighboring'' pairs $(c'_1,c'_2)$ of $(c_1,c_2)$,
where two pairs are neighbors if they differ in exactly one
co-ordinate, and by a distance of at most ten locations.
Clearly, here we do not touch upon the finer issues 
of efficient model searching, as these would
possibly require more sophisticated MCMC algorithms.

After a preliminary, exploratory simulation stage, we identified the
three {\em a posteriori} most likely pairs of thresholds as being
$(c_1^1,c_2^1)=(13.63,\, 2.72)$, $(c_1^2,c_2^2)=(13.89,\, 2.72)$,
$(c_1^3,c_2^3)=(13.78,\, 2.72)$. To estimate the actual
posterior probability of the most likely model,
$\pi(c_1^1,c_2^1|r)$, we define
$G_j(c_1,c_2)=\IND_{\{(c_1,c_2)=(c_1^j,c_2^j)\}}$, for $j=1,2,3$,
we set $F=G_1$, and we use the control variate $U=G-PG$ based
on $G=(G_1,G_2,G_3)^t$.
Writing $(c_1,c_2)\sim(c_1',c_2')$ when $(c_1,c_2)$ and
$(c_1',c_2')$ are neighboring pairs, $PG_j$ can be expressed,
for $j=1,2,3$, as,
$$PG_j(c_1,c_2)=
\begin{cases}
1-\frac{1}{40}\sum_{(c'_1,c'_2)\sim(c_1,c_2)}
    \min\Big\{1, \frac{p(r|c'_1,c'_2)}{p(r|c_1,c_2)}\Big\},
    & \text{if $(c_1,c_2)=(c_1^j,c_2^j)$;}\\
\frac{1}{40}
    \min\Big\{1, \frac{p(r|c^j_1,c^j_2)}{p(r|c_1,c_2)}\Big\},
    & \text{if $(c_1,c_2)\sim (c_1^j,c_2^j)$;}\\
0,
    & \text{otherwise.}\\
\end{cases}
$$

The resulting variance reduction factors 
obtained by $\mu_{n,\K}(F)$,
estimated from $T=100$ repetitions,
are {\bf 125.16, 32.83, 36.76, 30.90} and  {\bf 30.11},
after $n=10000, 20000,50000,100000$ and $200000$ 
simulation steps, respectively.
Figure \ref{f:finance} shows a typical simulation run.
All MCMC chains were initiated at $(c_1^1, c_2^1)$.

\begin{figure}[ht!]
\centerline{\includegraphics[width=4.1in]{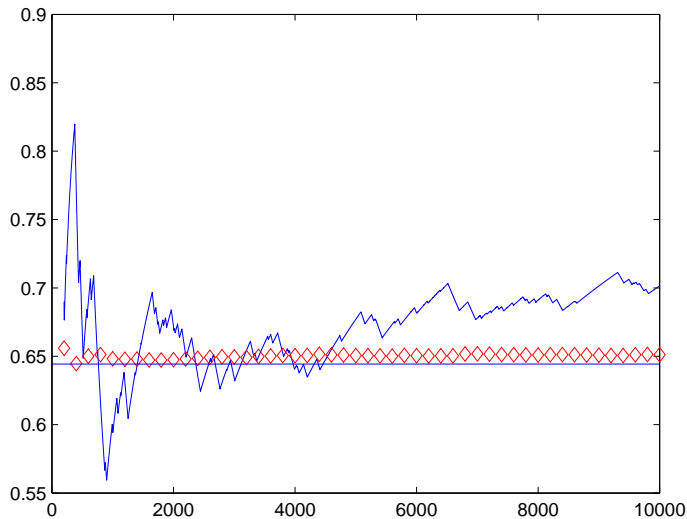}}
\caption{Threshold autoregressive model:
The sequence of the standard ergodic averages $\mu_n(F)$ is
shown as a solid blue line and the adaptive estimates
$\mu_{n,\K}(F)$, reported every $500$ iterations, as red
diamonds. The straight horizontal line represents the posterior
model probability obtained after $50$ million iterations.}
\flabel{finance}
\end{figure}


\subsection{A log-linear model}
We consider the $2 \times 3 \times 4$ table presented by
\citet{knuiman}, where $491$ subjects were classified according to
hypertension (yes, no), obesity (low, average, high) and alcohol
consumption (0, 1-2, 3-5, or $6+$ drinks per day).  We choose to
estimate the parameters of the log-linear model with three main
effects and no interactions, specified as,
$$ y_i \sim \mbox{Poisson}(\mu_i),~~\log(\mu_i) = x^t_i \beta,
~~i=1,2\ldots,24,$$
where the $y_i$ denote the cell frequencies,
modeled as Poisson variables with corresponding means
$\mu_i$, each $x_i$ is the $i$th row of the
$24 \times 7$ design matrix $x$, based on sum-to-zero constraints,
and $\beta=(\beta_1,\beta_2,\ldots,\beta_7)^t$ is the parameter vector.
In \citet{D-forster} this model was identified as having
the highest posterior probability among all log-linear interaction
models, under various prior specifications.

Assuming a flat prior on $\beta$,
standard Bayesian inference via MCMC can
be performed either by a Gibbs sampler that exploits
the log-concavity of full conditional
densities as in \citet{Dell}, or by a multivariate, random walk
Metropolis-Hastings sampler, in which an initial maximum likelihood
estimate of the covariance matrix gives guidance as to
the form of the proposal density.  Instead, here
we use a simple random-scan Gibbs sampler,
noting that a sample from the full conditional
density of each $\beta_j$
can be obtained directly as the logarithm
of a Gamma random variable with density,
\be
\mbox{Gamma} \left( \textstyle{\sum_i} y_i x_{ij},
\sum_{i:x_{ij}=1} \exp \Big\{ \sum_{\ell \neq
j} \beta_\ell x_{i\ell}
\Big\} \right).
\label{eq:logG}
\ee

In order to estimate the posterior mean of all seven
components of $\beta$,
we set $F_j(\beta)=\beta_j$ for all $j$,
and we use the same seven control variates
$U_1,U_2,\ldots,U_7$
for each $F_j$, where the
$U_\ell=G_\ell-PG_\ell$ are defined
in terms of the functions,
$G_{\ell}(\beta) = \exp(\beta_\ell)$,
$\ell=1,2,\ldots,7$.  The computation
of $PG_\ell$ is straightforward
since, in view of (\ref{eq:logG}),
the mean of $\exp(\beta_j)$
under the full conditional density
of $\beta_j$ is,
$$
\frac{\sum_i y_i x_{ij} } {
\sum_{i:x_{ij}=1} \exp \left( \sum_{\ell \neq j} \beta_\ell x_{i\ell}
\right)  }.$$

The variance reduction factors obtained
by $\mu_{n,\K}(F_j)$ after $n=100000$
simulation steps range between 
57.16 and 170.34, for different parameters
$\beta_j$. More precisely, averaging over $T=100$
repetitions, the variance reduction factors obtained
by $\mu_{n,\K}(F_j)$ are in the range,
{\bf 3.55--5.57, 38.2--57.69, 66.20--135.51,
57.16--170.34} and {\bf 85.41--179.11}, 
after $n=1000, 10000,50000,$ $100000$ and $200000$ 
simulation steps, respectively.
Figure~\ref{f:logl} shows an example 
of a sequence of ergodic averages for $\beta_7$.
All MCMC chains were initiated from the
corresponding maximum likelihood estimates.

\begin{figure}[ht!]
\centerline{\includegraphics[width=4.1in]{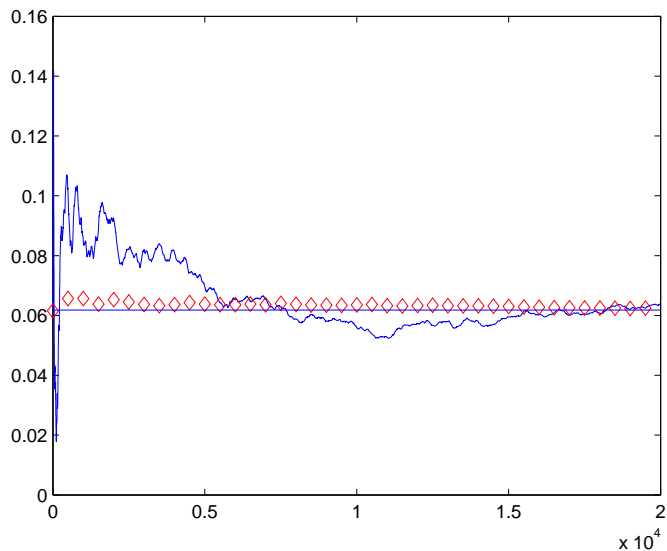}}
\caption{Log-linear model: The sequence of the standard ergodic
averages $\mu_n(F_7)$ for $\beta_7$ is shown as a solid blue line 
and the adaptive estimates $\mu_{n,\K}(F_7)$, plotted every $500$,
iterations, as red diamonds. The straight horizontal line represents
the estimate obtained after $5$ million iterations.}
\flabel{logl}
\end{figure}



\section{Theory}
\slabel{theory}

In this section we give precise conditions under which the
asymptotics developed in Sections~\ref{s:cv}
and~\ref{s:multiple} are rigorously
justified. The results together with their
detailed assumptions are stated below
and the proofs are contained in the appendix.
Note that, since the first two estimators we considered,
$\mu_{n,\trev,1}(F)$ and
$\mu_{n,\trev,2}(F)$, are special cases
of the estimators $\mu_{n,\Gamma}(F)$ and
$\mu_{n,\K}(F)$ introduced in \Section{multiple},
here we concentrate on the more general
estimators $\mu_{n,\Gamma}(F)$,
$\mu_{n,\K}(F)$.

First we recall the basic setting from \Section{cv}.
We take $\{X_n\}$ to be a Markov chain with values in
a general measurable space $\state$ equipped with a
$\sigma$-algebra $\clB$. The distribution of $\{X_n\}$
is described by its initial state $X_0=x\in\state$
and its transition kernel, $P(x,dy)$, as in (\ref{eq:kernel}).
The kernel $P$, as well as any of its powers $P^n$,
acts linearly on functions $F:\state\to\RL$ via,
$PF(x)=E[F(X_1)|X_0=x]$.

Our first assumption on the chain $\{X_n\}$ is that
\textit{$\psi$-irreducible} and \textit{aperiodic}.
This means that there is a $\sigma$-finite measure
$\psi$ on $(\state,\clB)$ such that,
for any $A\in\clB$ satisfying $\psi(A)>0$
and any initial condition $x$,
\[
P^n(x,A) > 0,\qquad \hbox{for all $n$ sufficiently large.}
\]
Without loss of generality, $\psi$ is assumed to
be \textit{maximal} in the sense that any other
such $\psi'$ is absolutely continuous with respect
to $\psi$.

Our second, and stronger, assumption, is an essentially minimal
ergodicity condition; cf. \citet{meyn-tweedie:book}: We assume that
there are functions $V:\state\to [0,\infty)$, $W:\state\to
[1,\infty)$, a ``small'' set $C\in\clB$, and a finite constant $b>0$
such that the Lyapunov drift condition (V3) holds:
$$PV-V\leq - W+b\IND_C.
\eqno{\hbox{(V3)}}
$$
Recall that a set $C\in\clB$ is small if there exists an integer
$m\geq 1$, a $\delta>0$ and a probability measure $\nu$ on
$(\state,\clB)$ such that,
$$P^m(x,B)\geq\delta\nu(B)\;\;\;\;\mbox{for all }\;
x\in C,\;B\in\clB.$$
Under (V3), we are assured that the chain
is positive recurrent, and that it possesses
a unique invariant (probability) measure
$\pi$. Our final assumption on the chain is that the
Lyapunov function $V$ in (V3) satisfies,
$\pi(V^2)<\infty$.

These assumptions are summarized as follows:
$$
\left.
\mbox{\parbox{.73\hsize}{\raggedright
    The chain $\{X_n\}$ is $\psi$-irreducible
    and aperiodic, with unique invariant measure
    $\pi$, and there exist functions
    $V:\state\to [0,\infty)$, $W:\state\to [1,\infty)$,
    a small set $C\in\clB$, and a finite constant
    $b>0$, such that (V3) holds and $\pi(V^2)<\infty$.
    }}
\;\;\right\}
\eqno{\mbox{(A)}}
$$
Although these conditions may seem somewhat involved,
their verification is generally straightforward;
see the texts \citet{meyn-tweedie:book,robert:book},
as well as some of the examples developed
in
\citet{roberts-tweedie:96,hobert-geyer:98,jarner-hansen:00,%
fort-et-al:03,roberts-rosenthal:04}. In fact, it is often possible
to avoid having to verify (V3) directly, by appealing to the
property of geometric ergodicity, which is essentially equivalent to
the requirement that (V3) holds with $W$ being a multiple of the
Lyapunov function $V$. For large classes of MCMC samplers, geometric
ergodicity has been established in the above papers, among others.
Moreover, geometrically ergodic chains, especially in the reversible
case, have many attractive properties, as discussed, for example, by
\citet{roberts-rosenthal:98}.

In the interest of generality, the main
results of this section are stated in terms of the weaker
(and essentially minimal) assumptions in~(A).
Some details on general strategies for their
verification can be found in the references above.

Apart from conditions on the Markov chain $\{X_n\}$,
the asymptotic results stated earlier also require
some assumptions on the function $F:\state\to\RL$
whose mean under $\pi$ is to be estimated,
and on the (possibly vector-valued) function
$G:\state\to\RL^k$ which is used for the
control variate $U=G-PG$. These assumptions
are most conveniently stated within the
weighted-$L_\infty$ framework of
\citet{meyn-tweedie:book}.
Given an arbitrary function
$W:\state\to[1,\infty)$, the weighted-$L_\infty$
space $L_\infty^W$ is the Banach space,
$$L_\infty^W:=\Big\{\mbox{functions}\;F:\state\to \RL\;\;
\mbox{s.t.}\;\;\|F\|_W:=\sup_{x\in\state}
\frac{|F(x)|}{W(x)}<\infty\Big\}.$$
With a slight abuse of notation, we
say that a vector-valued function
$G=(G_1,G_2,\ldots,G_k)^t$ is in $L_\infty^W$
if $F_j\in L_\infty^W$ for each $j$.

\medskip

\noindent
{\bf Theorem 1. }
{\em
Suppose the chain $\{X_n\}$
satisfies conditions {\em (A)}, and let
$\{\theta_n\}$ be any
sequence of random vectors in $\RL^k$
such that $\theta_n$ converge to some
constant $\theta\in\RL^k$
a.s., as $n\to\infty$.
Then:
\begin{enumerate}
\item[{\bf (i)}]
{\sc [Ergodicity] }
The chain is positive Harris recurrent, it  has a unique
invariant (probability) measure $\pi$, and it
converges in distribution to $\pi$,
in that for any $x\in\state$ and $A\in\clB$,
$$P^n(x,A)\to\pi(A),\;\;\;\;\mbox{as }\;n\to\infty.$$
In fact, there
exists a finite constant $B$ such that,
\be
\sum_{n=0}^\infty |P^nF(x)-\pi(F)|\leq B(V(x)+1),
\label{eq:V3bound}
\ee
uniformly over all initial states $x\in\state$
and all function $F$ such that $|F|\leq W$.
\item[{\bf (ii)}]
{\sc [LLN] }
For any $F,G\in L_\infty^W$ and any $\vartheta\in\RL^k$,
write $U=G-PG$ and $F_\vartheta:=F-\langle\vartheta,U\rangle$.
Then
the ergodic averages $\mu_n(F)$,
as well as the
adaptive averages
$\mu_n(F_{\theta_n})$,
both converge to $\pi(F)$ a.s.,
as $n\to\infty$.
\item[{\bf (iii)}]
{\sc [Poisson Equation] }
If $F\in L_\infty^W$, then there exists
a solution $\hat{F}\in L_\infty^{V+1}$ to
the Poisson equation, $P\hat{F}-\hat{F}=-F+\pi(F)$,
and $\hat{F}$ is unique up to an additive constant.
\item[{\bf (iv)}]
{\sc [CLT for $\mu_n(F)$] }
If $F\in L_\infty^W$ and the variance,
$\sigma_{F}^2:=\pi(\hat{F}^2-(P\hat{F})^2)$ is nonzero,
then the normalized ergodic averages
$\sqrt{n}[\mu_n(F)-\pi(F)]$
converge in distribution to $N(0,\sigma_F^2)$,
as $n\to\infty$.
\item[{\bf (v)}]
{\sc [CLT for $\mu_n(F_{\theta_n})$] }
If $F,G\in L_\infty^W$,
and the variances,
$\sigma_{F_{\theta}}^2:= \pi(\hat{F}_{\theta}^2-(P\hat{F}_{\theta})^2)$
and
$\sigma_{U_j}^2:= \pi(\hat{U}_j^2-(P\hat{U}_j)^2)$,
$j=1,2,\ldots,k$ are all
nonzero, then the normalized adaptive
averages $\sqrt{n}[\mu_n(F_{\theta_n})-\pi(F)]$
converge in distribution to $N(0,\sigma_{F_{\theta}}^2)$,
as $n\to\infty$.
\end{enumerate}
}

\medskip

Suppose the chain $\{X_n\}$ satisfies conditions
(A) above, and that the functions $F$
and $G=(G_1,G_2,\ldots,G_k)^t$ are
in $L_\infty^W$. Theorem~1 states that
the ergodic averages $\mu_n(F)$ as well
as the modified averages
$\mu_n(F_\theta)$ based on the
vector of control variates
$U=G-PG$ both converge to $\pi(F)$,
and both are asymptotically Normal.
Next we examine the choice of
the parameter vector $\theta=\theta^*$
which minimizes the limiting variance
$\sigma_{F_\theta}^2$
of the modified averages, and the
asymptotic behavior of
the estimators $\hat{\theta}_{n,\Gamma}$
and $\hat{\theta}_{n,\K}$ for $\theta^*$.

As in \Section{multiple}, let
$\Gamma(G)$ denote the $k\times k$
matrix with entries,
$\Gamma(G)_{ij}=\pi(G_iG_j-(PG_i)(PG_j))$,
and recall that, according to
Theorem~1, there exists a solution $\hat{F}$
to the Poisson equation for $F$. The simple
computation outlined in \Section{multiple}
(and justified in the proof of Theorem 2)
leading to equation (\ref{eq:thetaStar-k})
shows that the variance $\sigma_{F_\theta}^2$
is minimized by the choice,
\ben
\theta^*=
\Gamma(G)^{-1}\pi(\hat{F}G-(P\hat{F})(PG)),
\een
as long as the matrix $\Gamma(G)$ is invertible.
Our next result establishes the a.s. consistency
of the estimators,
\ben
\hat{\theta}_{n,\Gamma}
&=&
    \Gamma_n(G)^{-1}
    [\mu_n(F(G+PG))-
    \mu_n(F)\mu_n(G+PG)]\\
\;\;\;\;
\hat{\theta}_{n,\K}
&=&
    \K_n(G)^{-1}
    [\mu_n(F(G+PG))-
    \mu_n(F)\mu_n(G+PG)],
\een
where the $k\times k$
matrices $\Gamma_n(G)$ and $\K_n(G)$ are
defined, respectively, by,
\ben
(\Gamma_n(G))_{ij}
&=&
    \mu_n(G_iG_j)-\mu_n((PG_i)(PG_j))\\
\mbox{and}
\;\;\;\;
(\K_n(G))_{ij}
&=&
\frac{1}{n}\sum_{t=0}^{n-1}
(G_i(X_t)-PG_i(X_{t-1})) (G_j(X_t)-PG_j(X_{t-1})).
\een

\medskip

\noindent
{\bf Theorem 2. }
{\em
Suppose that the chain $\{X_n\}$ is reversible and
satisfies conditions {\em (A)}. If the functions
$F,G$ are both in $L_\infty^W$
and the matrix $\Gamma(G)$ is nonsingular,
then both the adaptive estimators for
$\theta^*$ are a.s. consistent:
\ben
\hat{\theta}_{n,\Gamma}\to\theta^*
&&
    \;\;\;\mbox{a.s., as }\;n\to\infty;\\
\hat{\theta}_{n,\K}\to\theta^*
&&
    \;\;\;\mbox{a.s., as }\;n\to\infty.
\een
}

\medskip

Recall the definitions of the two estimators
$\mu_{n,\Gamma}(F)$ and $\mu_{n,\K}(F)$
from equations (\ref{eq:multi-est1})
and (\ref{eq:multi-est2}) in \Section{multiple}.
Combining the two theorems, yields the desired
asymptotic properties of the two estimators:

\medskip

\noindent
{\bf Corollary 1. }
{\em Suppose that the chain $\{X_n\}$ is reversible
and satisfies conditions {\em (A)}. If the functions
$F,G$ are both in $L_\infty^W$
and the matrix $\Gamma(G)$ is nonsingular,
then the adaptive estimators
$\mu_{n,\Gamma}(F)$
and $\mu_{n,\K}(F)$
for $\pi(F)$ satisfy:
\begin{enumerate}
\item[{\bf (i)}]
{\sc [LLN] }
The adaptive estimators
$\mu_{n,\Gamma}(F)$,
$\mu_{n,\K}(F)$
both converge to $\pi(F)$ a.s.,
as $n\to\infty$.
\item[{\bf (ii)}]
{\sc [CLT] } If $\sigma_{F_{\theta^*}}^2:=
\pi(\hat{F}_{\theta^*}^2-(P\hat{F}_{\theta^*})^2)$ is nonzero, then
the normalized adaptive averages
$\sqrt{n}[\mu_{n,\Gamma}(F)-\pi(F)]$ and
$\sqrt{n}[\mu_{n,\K}(F)-\pi(F)]$ converge in distribution to
$N(0,\sigma_{F_{\theta^*}}^2)$, as $n\to\infty$, where the variance
$\sigma_{\theta^*}^2$ is minimal among all estimators based on the
control variate $U=G-PG$, in that
$\sigma_{\theta^*}^2=\min_{\theta\in\RL^k} \sigma_{\theta}^2.$
\end{enumerate}
}

\medskip

Some additional results on the long-term behavior of estimators
similar to the ones considered above can be found in Meyn's recent
work in \citet{meyn:06} and \citet[Chapter~11]{meyn:book}.


\section{Extensions}

We have presented a series of small and large 
simulation experiments motivated by important 
classes of Bayesian inference problems,
and we have repeatedly observed that generally 
straightforward choices for functions $G$
in the construction of control variates $U=G-PG$
provide very effective variance reduction
results. Moreover, the methodology utilizing
these control variates can be implemented 
as an essentially black-box,
post-processing algorithm.

The theory presented is applicable to any reversible 
Markov chain. Our focus here has been primarily
on cases of samplers for which we can find
{\em some} functions $G$ such that
the one-step conditional expectations 
required for the computation of $PG$ are available 
in closed form. These are readily available in all 
conjugate Gibbs and discrete Metropolis algorithms,
as well as in most Markovian models for stochastic
networks; see \citet{meyn:book} and the references
therein.

Most of our experiments were performed using random-scan 
Gibbs samplers, in order to maintain reversibility; 
this is not necessarily a restrictive choice, 
since the convergence properties of random-scan algorithms 
are comparable to those of systematic-scan samplers; see
\citet{robsah97}.  Moreover, any implementation technique
that can facilitate or speed up the MCMC convergence 
(such as blocking schemes, transformations, 
other reversible chains, and so on), can be 
used, as long as reversibility is maintained.

There are many other Gibbs sampling algorithms in which full
conditional density expectations are analytically available and,
therefore, our proposed methodology is immediately applicable. 
Apart from the natural extensions of the examples in
Section~\ref{sec:simple}, we emphasize that conjugate Gibbs 
sampling is the key ingredient in Bayesian inference 
for dynamic linear models, see
\citet{reis}; slice Gibbs auxiliary variables applications, see
\citet{damien99}; Dirichlet processes, see \citet{MacEachernmuller};
and spatial regression models, see \citet{gamerman03}.

Metropolis-Hastings algorithms were used in
Example~5 and in the two-threshold autoregressive
model in Section~\ref{s:setar}. In both cases,
the samplers operate on a discrete state space,
making it possible to compute $PG$ 
directly in closed form.
It may be worth emphasizing that for
any discrete Metropolis-Hastings sampler where
the number of possible proposed moves is not 
prohibitively large, the function $PG$ can be
easily analytically obtained for any choice of $G$.

In closing, we note that the main obstacle in the 
immediate applicability of our methodology is the
presence of the accept/reject probability in
Metropolis-Hastings steps with continuous 
proposals. The ways in which this methodology 
can be applied in such cases
are explored in ongoing work that investigates 
this issue in detail, and which will be reported 
in \citet{DKMT:prep}.


\section*{Acknowledgments}
We are grateful to Sean Meyn and Zoi Tsourti for many interesting
conversations related to this work.

\appendix
\section*{Appendix: Proofs of Theorems 1 and 2}

\noindent
{\sc Proof of Theorem 1. } Since any small set is
petite, \citet[Section~5.5.2]{meyn-tweedie:book},
the $f$-norm ergodic theorem of
\citet{meyn-tweedie:book} implies that
$\{X_n\}$ is positive recurrent with
a unique invariant measure $\pi$
such that (\ref{eq:V3bound}) holds,
and \citet[Theorem~11.3.4]{meyn-tweedie:book}
proves the Harris property, giving~(i).

From \citet[Theorem~14.0.1]{meyn-tweedie:book}
we have that, under (V3), $\pi(W)<\infty$.
Since $F$ is in $L_\infty^W$, $\pi(|F|)$ is finite,
and since $G\in L_\infty^W,$ Jensen's inequality
guarantees that $\pi(|U|)$ is finite. The invariance
of $\pi$ then implies that $\pi(U)=0$;
therefore,
\citet[Theorem~17.0.1]{meyn-tweedie:book}
shows that $\mu_n(F)\to\pi(F)$ and $\mu_n(U)\to 0$
a.s.\ as $n\to\infty$, and since
$\theta_n\to\theta$ by assumption,
$\mu_n(F_\theta)$ also converges
to $\pi(F)$ a.s., proving~(ii).

The existence of a solution $\hat{F}$ to
the Poisson equation in~(iii) follows from
\citet[Theorem~17.4.2]{meyn-tweedie:book},
and its uniqueness from
\citet[Theorem~17.4.1]{meyn-tweedie:book}.
The CLT in~(iv) is a consequence of
\citet[Theorem~17.4.4]{meyn-tweedie:book}.

Finally, since $F,G\in L_\infty^W$,
the functions $U$ and $F_\theta$ are in
$L_\infty^W$ too, so $\hat{U}_j$ and
$\hat{F}_\theta$ exist for each $j=1,2,\ldots,k$.
As in~(iv), the scaled averages
$\sqrt{n}[\mu_n(F_\theta)-\pi(F)]$
and $\sqrt{n}\mu_n(U_j)$
converge in distribution to
$N(0,\sigma_{F_\theta}^2)$ and
$N(0,\sigma_{U_j}^2)$,
respectively, for each $j$,
where the variances
$\sigma_{F_\theta}^2$
and $\sigma_{U_j}^2$
are as in~(iii).
Writing $\theta=(\theta_1,\theta_2,\ldots,\theta_k)^t$
and
$\theta_n=(\theta_{n,1},\theta_{n,2},\ldots,\theta_{n,k})^t$,
we can express,
$$\sqrt{n}[\mu_n(F_{\theta_n})-\pi(F)]
= \sqrt{n}[\mu_n(F_\theta)-\pi(F)]+
\sum_{j=1}^k\Big\{
(\theta_{n,j}-\theta_j)
\sqrt{n}\mu_n(U_j)
\Big\}.$$
Each of the terms in the
second sum on the right-hand-side
above converges to zero in probability,
since and $\sqrt{n}\mu_n(U_j)$ converges
to a Normal distribution and
$(\theta_{n,j}-\theta_j)\to 0$ a.s.
Therefore, the sum converges to zero
in probability, and the CLT in~(v)
follows from (iv).
\qed

\medskip

Note that the assumption $\sigma_{U_j}^2\neq 0$ in
the theorem is not necessary, since the case
$\sigma_{U_j}^2=0$ is trivial in view
of \citet[Proposition~2.4]{kontoyiannis-meyn:Itmp},
which implies that, then,
$\sqrt{n}\mu_n(U_j)\to 0$ in probability,
as $n\to\infty$.

\medskip

\noindent
{\sc Proof of Theorem 2. }
We begin by justifying the computations in
\Section{multiple}.
Define
$\sigma_{F_\theta}^2=\pi(\hat{F}_\theta^2- (P\hat{F}_\theta)^2)$,
where $\hat{F}$ exists by Theorem~1.
Since $\hat{F}$ solves the Poisson equation for $F$,
it is easy to check that $\hat{F}_\theta:=\hat{F}-\langle\theta,G\rangle$
solves the Poisson equation for $F_\theta$.
Substituting this in the above expression
for $\sigma_{F_\theta}^2$ yields (\ref{eq:varTheta}).
To see that all the functions in (\ref{eq:varTheta})
are indeed integrable recall that
$\hat{F}\in L_\infty^{V+1}$
and note that,
since $V$ is nonnegative, (V3) implies that
$1\leq W\leq V + b\IND_C$, hence
$\pi(W^2)$ is finite since $\pi(V^2)$
is finite by assumption.
Therefore, since $G\in L_\infty^{W}$,
$\hat{F}$ and $G$ are both in $L_2(\pi)$,
and H\"{o}lder's inequality implies that
$\pi(\hat{F}\langle\theta,G\rangle)$
is finite. Finally, Jensen's inequality
implies that $P\hat{F}$ and $PG$ are also
in $L_2(\pi)$, so that $\pi(P\hat{F}\langle\theta,PG\rangle)<\infty$.
And, for the same reasons, all the functions
appearing in the computations leading to the results
of Lemma~2 and Proposition~2 are also integrable.

The expression for the
optimal $\theta^*$ in (\ref{eq:thetaStar-k})
is simply the solution for the minimum of the
quadratic in (\ref{eq:varTheta}). Again,
note that $\hat{F},G,P\hat{F}$ and $PG$
are all in $L_2(\pi)$ so $\theta^*$
is well-defined.

The consistency proofs follow from repeated
applications of the ergodic theorems established
in Theorem~1. First note that, since
$G\in L_\infty^W$ and $\pi(W^2)<\infty$ as
remarked above, the product $G_iG_j$ is
$\pi$-integrable, and by Jensen's inequality
so is any product
of the form $(PG_i)(PG_j)$. Therefore,
the ergodic theorem of \citet[Theorem~17.0.1]{meyn-tweedie:book}
implies that $\Gamma_n(G)\to\Gamma(G)$ a.s.
Similarly, the functions $F$, $G$, $PG$, $FG$ and $FPG$
are all $\pi$-integrable, so that the same ergodic
theorem implies that $\hat{\theta}_{n,\Gamma}$
indeed converges to $\theta^*$ a.s., as $n\to\infty$.

To establish the corresponding result for
$\hat{\theta}_{n,\K}$, it suffices to show that
$K_n(G)\to K(G)$ a.s., and to that end we consider
the bivariate chain $Y_n=(X_n,X_{n+1})$ on the
state space $\state\times\state$. Since
$\{X_n\}$ is $\psi$-irreducible and aperiodic,
$\{Y_n\}$ is $\psi^{(2)}$-irreducible and aperiodic
with respect to the bivariate measure
$\psi^{(2)}(dx,dx'):=\psi(dx)P(x,dy)$.
Given functions $W,V$ a small set $C$ and
a constant $b$ so that (V3) holds, it is
immediate that (V3) also holds for $\{Y_n\}$
with respect to the functions
$V^{(2)}(x,x')=V(x')$,
$W^{(2)}(x,x')=W(x')$, the small
set $\state\times C$, and the same $b$.
The unique invariant measure of
$\{Y_n\}$ is then
$\pi^{(2)}(dx,dx'):=\pi(dx)P(x,dy)$,
and $\pi^{(2)}(V^{(2)})$ is finite.
Therefore, assumptions (A) hold for
$\{X_n\}$ and, for each pair
$1\leq i,j\leq k$ we can invoke the
ergodic theorem
\citet[Theorem~17.0.1]{meyn-tweedie:book}
for the $\pi^{(2)}$-integrable function,
$$H(x,x'):=
(G_i(x')-PG_i(x)) (G_j(x')-PG_j(x)),
$$
to obtain that, indeed, $\K_n(G)\to \K(G)$ a.s.
\qed

\medskip

\noindent
{\sc Proof of Corollary~1. }
The ergodic theorems in~(i) are immediate
consequences of Theorem~1~(ii) combined
with Theorem~2. The computation
in \Section{multiple} which shows that
$\theta^*$ in (\ref{eq:thetaStar-k})
indeed minimizes $\sigma_{F_\theta}^2$
(justified in the proof of Theorem~2)
shows that $\sigma_{\theta^*}^2=\min_{\theta\in\RL^k}
\sigma_{\theta}^2.$
Finally, the assumption that $\Gamma(G)$
is nonsingular combined with Lemma~2,
imply that
all the variances $\sigma_{U_j}^2$ must be
nonzero. Therefore, Theorem~2 combined
with the central limit theorems
in parts~(iv) and~(v) of Theorem~1, prove
part~(ii) of the Corollary.
\qed

{\small
\bibliographystyle{chicago}

}

\end{document}